# Reentrant polar phase induced by the ferro-ionic coupling in $Bi_{1-x}Sm_xFeO_3$ nanoparticles

Anna N. Morozovska[1*], Eugene A. Eliseev[2], Igor V. Fesych[3], Yuriy O. Zagorodniy[2], Oleksandr S. Pylypchuk[1], Evgenii V. Leonenko[1], Maxim V. Rallev[1], Andrii D. Yaremkevych[1], Lesya P. Yurchenko[2], Lesya Demchenko[4], Sergei V. Kalinin[5†], and Olena M. Fesenko[1‡]

[1] Institute of Physics, National Academy of Sciences of Ukraine,46, pr. Nauky, 03028 Kyiv, Ukraine

[2]Frantsevich Institute for Problems in Materials Science, National Academy of Sciences of Ukraine, 3, str. Omeliana Pritsaka, 03142 Kyiv, Ukraine

[3]Taras Shevchenko National University of Kyiv, 01030 Kyiv, Ukraine

[4]Stockholm University, Department of Materials and Environmental Chemistry, Sweden and Ye. O. Paton Institute of Materials Science and Welding, National Technical University of Ukraine "Igor Sikorsky Kyiv Polytechnic Institute"

[5]Department of Materials Science and Engineering, University of Tennessee, Knoxville, TN, 37996, USA

Using the model of four sublattices, the Landau-Ginzburg-Devonshire-Kittel phenomenological approach and the Stephenson-Highland ionic adsorption model for the description of coupled polar and antipolar long-range orders in ferroics, we calculated analytically the phase diagrams and polar properties of $Bi_{1-x}Sm_xFeO_3$ nanoparticles covered by surface ions in dependence on their size, surface ions density, samarium content "x" and temperature. The size effects and ferro-ionic coupling govern the appearance and stability conditions of the long-range ordered ferroelectric, reentrant ferrielectric and antiferroelectric phases in the $Bi_{1-x}Sm_xFeO_3$ nanoparticles. Calculated phase diagrams are in a qualitative agreement with the X-ray diffraction phase analysis, electron paramagnetic resonance, infra-red spectroscopy and electrophysical measurements of the $Bi_{1-x}Sm_xFeO_3$ nanopowders sintered by the solution combustion method. The combined theoretical-experimental approach allows to explain the influence of the ferro-ionic coupling and size effects in $Bi_{1-x}Sm_xFeO_3$ nanoparticles on their polar properties.

* corresponding author, e-mail: anna.n.morozovska@gmail.com
† corresponding author, e-mail: sergei2@utk.edu
‡ corresponding author, e-mail: fesenko.olena@gmail.com

## I. INTRODUCTION

Due to the strong influence of surface and size effects nanoscale multiferroics of various shape and sizes are unique model objects for fundamental studies of the polar, antipolar and/or magnetic long-range ordering and magnetoelectric coupling [1]. The ferroelectric-antiferroelectric-antiferromagnetic solid solution $Bi_{1-x}Sm_xFeO_3$ is a classical multiferroic, whose thin films have a large spontaneous polarization, a high magnetoelectric coupling coefficient and conductive domain walls [2, 3, 4, 5]. The polar, magnetic and multiferroic properties of $Bi_{1-x}Sm_xFeO_3$ thin films and multilayer structures are relatively well studied experimentally and theoretically [6, 7, 8, 9].

However, the properties of $Bi_{1-x}Sm_xFeO_3$ nanoparticles [10, 11] are much less studied experimentally and almost unexplored theoretically. Due to the surface screening and size effects one can expect that $Bi_{1-x}Sm_xFeO_3$ nanoparticles can be stabilized in the unusual phase states, such as the mixed and/or reentrant ferrielectric phases, morphotropic phase boundaries and spatially-modulated incommensurate phases inherent to the rare-earth doped orthoferrites and tilted perovskites [2-4]. Also, one can suppose that the $Bi_{1-x}Sm_xFeO_3$ nanoparticles, similarly to other ferroelectric nanoparticles, can be very promising fillers for the energy storage and harvesting [12, 13, 14], and multi-bit memories [15, 16]. Thus, the analytical description of the phase diagrams, long-range ordering and magnetoelectric coupling in $Bi_{1-x}Sm_xFeO_3$ nanoparticles seems urgent to explain, predict and optimize their polar, dielectric, magnetic and magnetoelectric properties.

The theoretical model of four sublattices $A_i$ ($i = 1 - 4$), shortly FSM [17], allows the analytical description of cation displacement in the $Bi_{1-x}Sm_xFeO_3$, where the samarium content "$x$" and temperature $T$ determine the stability of the long-range ordered rhombohedral ferroelectric (FE), mixed ferrielectric (FEI), antiferroelectric (AFE), and paraelectric/dielectric nonpolar (NP) orthorhombic phases. Using the FSM model and the Landau-Ginzburg-Devonshire-Kittel (LGDK) phenomenological approach, the phase diagrams of the strained thin films of $Bi_{1-x}Sm_xFeO_3$ have been calculated in agreement with experiment [18]. The FSM-LGDK approach reduces the description of the phase diagram and polar properties of the bulk $Sm_xBi_{1-x}FeO_3$ to the thermodynamic analysis of the Landau-type free energy with several phenomenological parameters: sublattices linear stiffness, sublattices coupling strength and the gradient energy coefficients [19].

In the case of the $Bi_{1-x}Sm_xFeO_3$ nanoparticles suspension in different environments the Stephenson-Highland (SH) ionic adsorption can occur at the ferroelectric surface [20, 21]. Within the SH model the dependence of the surface charge density on electric potential excess at the surface of the nanoparticle is controlled by the concentration of positive and negative surface charges. A combination of the LGDK and SH approaches allows for the derivation of analytical solutions describing unusual ferro-ionic states [22, 23] in uniaxial [24, 25, 26] and multiaxial [27] ferroelectric thin films and nanoparticles [28], as well as antiferroelectric thin films with electrochemical polarization switching [29, 30]. The ferro-ionic states are the result of nonlinear electrostatic interaction between the ions with the surface charge density obeyed Langmuir adsorption isotherm and ferroelectric dipoles [22-30].

The ferro-ionic and/or antiferro-ionic states can originate in the $Bi_{1-x}Sm_xFeO_3$ nanoparticles due to the strong nonlinear dependence of the screening charge density on the surface electric potential, also named the electrochemical overpotential [24]. In this work we analyze the influence of size effects and ferro-ionic coupling on the appearance and stability conditions of the FE, FEI and AFE phases in the $Sm_xBi_{1-x}FeO_3$ nanoparticles and compare theoretical results with experiments.

## II. THEORETICAL MODELLING

### A. Analytical calculations of the phase diagram of a bulk $Sm_xBi_{1-x}FeO_3$

Far from the temperature of the antiferrodistortive transition, which is well above 1200 K for the Sm-doped $BiFeO_3$, the structural order parameter ($FeO_6$ octahedra tilt) changes very weakly and its influence on the polar and antipolar long-range ordering can be taken into account as the constant renormalization of corresponding coefficients in the Landau free energy [18, 19, 31]. For the case one can write the Landau-Kittel free energy of a homogeneous bulk $Bi_{1-x}Sm_xFeO_3$ in the following dimensionless form [18, 19]:

$$F_{bulk} = \frac{1}{2}\alpha(T,x)P^2 + \frac{1}{2}\eta(T,x)A^2 + \frac{1}{4}\beta_{ij}\left(P_i^2P_j^2 + A_i^2A_j^2\right) + \frac{1}{2}\xi_{ij}P_i^2A_j^2 - \vec{P}\vec{E}. \quad (1)$$

Here $\vec{P} = \frac{1}{2}\left(\vec{C}_1 + \vec{C}_2 + \vec{C}_3 + \vec{C}_4\right)$ and $\vec{A} = \frac{1}{2}\left(\vec{C}_1 - \vec{C}_2 + \vec{C}_3 - \vec{C}_4\right)$ are the dimensionless polar and antipolar order parameters, respectively. They are expressed via the cation displacements, $C_i$, in terms of the FSM; $\vec{E}$ is the dimensionless electric field component coupled to the polarization $\vec{P}$. The magnitudes of the three-component vectors $\vec{P}$ and $\vec{A}$ are $P^2 = P_1^2 + P_2^2 + P_3^2$ and $A^2 = A_1^2 + A_2^2 + A_3^2$. The linear stiffnesses of sublattices are $\alpha(T,x)$ and $\eta(T,x)$ are isotropic due to the cubic parent phase of Sm-doped $BiFeO_3$. The coefficients $\beta_{ij} = \beta\delta_{ij}$ and $\xi_{ij} = \xi\delta_{ij}$ in the isotropic approximation, where $\beta$ is the ferroelectric nonlinearity coefficient (positive for the 2-4 Landau expansion), $\xi$ is the sublattices coupling strength, and $\delta_{ij}$ is the Kronecker delta symbol. The gradient, elastic and flexoelectric energies are omitted for a homogenous stress-free and strain-free bulk $Sm_xBi_{1-x}FeO_3$. The equivalence of the sublattices can lead to the assumption $\beta = 1$ without loss of generality. The details of the free energy nondimensionalization are described in **Appendix A** [32].

The functional form of the temperature and Sm-content dependence of the dimensionless coefficients $\alpha(T,x)$ and $\eta(T,x)$ in Eq.(1) were chosen to reproduce the phase diagram of a bulk $Bi_{1-x}Sm_xFeO_3$ observed experimentally [2], namely [18, 19]:

$$\alpha(T,x)=\frac{T}{T_C}-\exp\left[-\left(\frac{x}{x_C}\right)^4\right], \tag{2a}$$

$$\eta(T,x)=\eta_0\left\{\frac{T}{T_A}-\exp\left[-\left(\frac{x}{x_A}\right)^2\right]\right\}. \tag{2b}$$

The functional form of Eq.(2) was selected using the grounds discussed in Refs.[18, 19], where the quantitative theoretical analysis of the phase diagram and spontaneous polarization temperature dependences have been done.

Here we would like to mention that, in accordance with Landau theory, the coefficients $\alpha(T,x)$ and $\eta(T,x)$ should be linear functions of the temperature $T$, which changes their sign at the characteristic temperatures. The coefficient $\alpha(T,0)$ changes the sign at the Curie temperature $T_C$ corresponding to the polar phonon mode softening, and $\eta(T,0)$ changes its sign at the Neel temperature $T_A$ corresponding to the antipolar mode softening. Since the increase of Sm content x deteriorates the polar and the antipolar orderings and induces the transition to the NP paraelectric phase for x>0.15 in the bulk $Bi_{1-x}Sm_xFeO_3$, the inequalities $\alpha(T,x)>0$ for $T>T_C$ and $\eta(T,x)>0$ for $T>T_A$ is valid for $0\leq x\leq 1$. The dimensionless parameter $\xi$ should be in the range $\left\{-\frac{1}{2},\infty\right\}$ for the stability of the free energy (1). For the best agreement with experimental results, we can vary six fitting parameters, $\eta_0$, $T_C$, $T_A$, $x_C$, $x_A$, and $\xi$.

Phase diagram of a bulk $Bi_{1-x}Sm_xFeO_3$, calculated using the LGDK free energy (1) and expressions (2) for the expansion coefficients, is shown in **Fig. 1(a)**. The values of fitting parameters, $T_C=1100$ K, $T_A=800$ K, $x_C=0.1$, $x_A=0.15$, $\eta_0=0.138$ and $\xi=0.121$, correspond to the best agreement with the experimental phase diagram shown in fig.1(a) in Ref.[2].

The FE phase in **Fig.1(a)** corresponds to the nonzero spontaneous polarization ($P_S\neq 0$) and zero spontaneous antipolar order ($A_S=0$); the mixed FEI phase corresponds to $P_S\neq 0$ and $A_S\neq 0$, the AFE phase corresponds to $A_S\neq 0$ and $P_S=0$, and the NP phase corresponds to $P_S=0$ and $A_S=0$ (see **Fig. 1(b)** and **1(c)**). The polarization $\vec{P}$ is readily measurable experimentally by application of the electric field $\vec{E}$, and the antipolar order $\vec{A}$ cannot be directly measured. However, the biquadratic coupling between $\vec{P}$ and $\vec{A}$, namely the term $\frac{1}{2}\xi P^2A^2$ in Eq.(1), changes the field dependence $\vec{P}(\vec{E})$ and the dielectric susceptibility tensor, $\chi_{ij}=\frac{\partial P_i}{\partial E_j}$, in the case $\vec{A}\neq 0$. Simple analytical expressions for the energies of different phases, $P_S$ and $A_S$ values, used for the calculation of the diagrams in **Figs. 1(a)-(c)**, are listed in **Table A1** [32].

It is seen from the **Figs. 1(a)-(c)** that the increase of Sm content $x$ above 0.1 at $T > 300^{o}$C leads to the second order phase transition from the FE to the NP phase. The increase of $x$ above (0.075 – 0.1) at $T < 300^{o}$C leads to the appearance of the mixed FEI phase, which transforms in the AFE phase with further increase in $x$. The further increase in $x$ leads to the gradual degradation and eventual disappearance of the antipolar order (e.g., for $x > 0.15$ at $0^{o}$C), and to the stability of the NP phase.

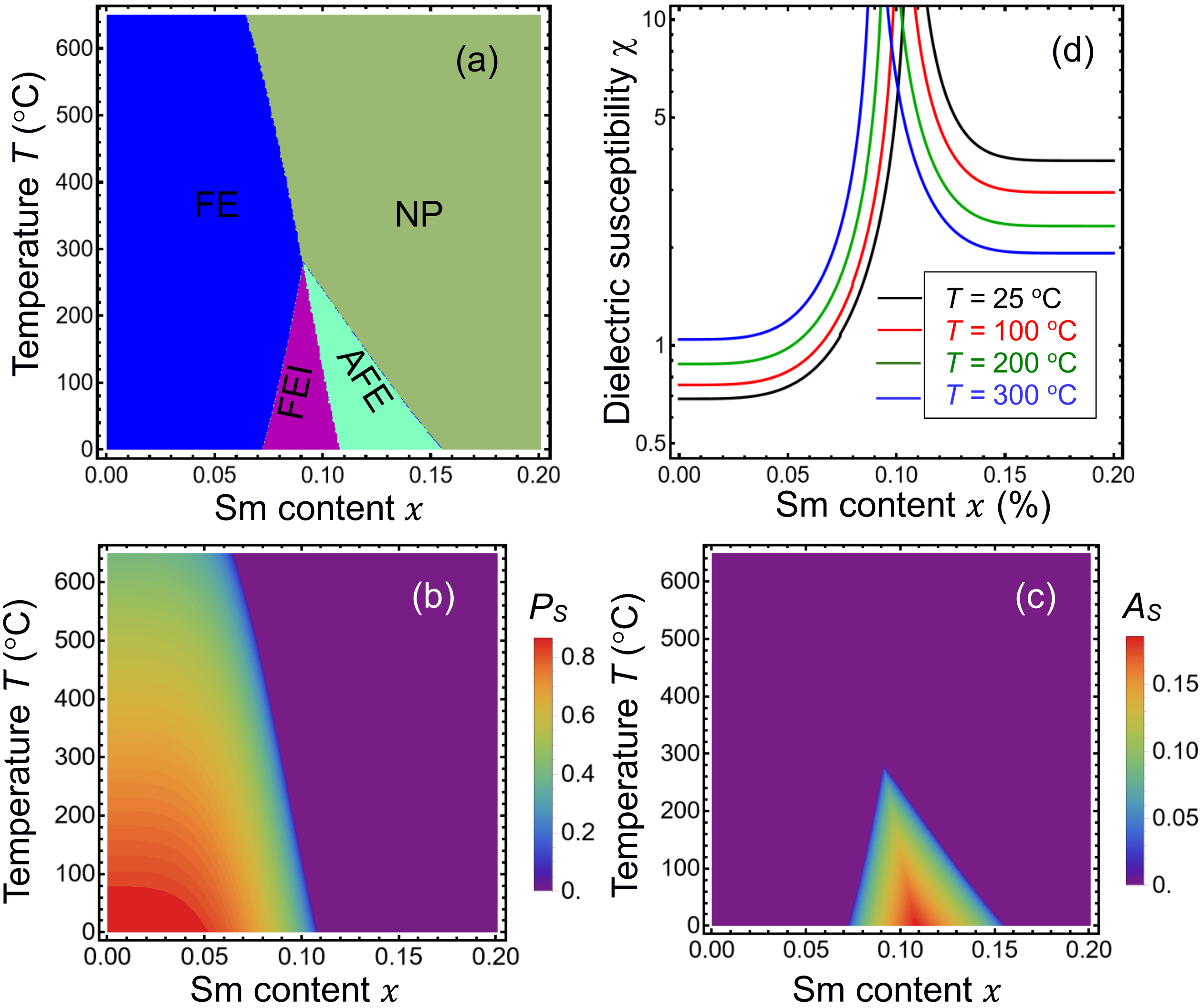

**FIGURE 1.** The phase diagram of a bulk $Sm_xBi_{1-x}FeO_3$ **(a)**, spontaneous polarization $P_S$ **(b)** and antipolar order parameter $A_S$ **(c)** in dependence on the temperature $T$ and Sm content $x$ calculated using the free energy (1) with the following parameters: $T_C = 1100$ K, $T_A = 800$ K, $x_C = 0.10$, $x_A = 0.15$, $\eta_0 = 0.138$, $\beta = 1$, and $\xi = 0.121$. **(d)** The dependence of the $Sm_xBi_{1-x}FeO_3$ linear dielectric susceptibility on the Sm

content $x$ calculated at $T$ =25°C (the black curve), 100°C (the red curve), 200°C (the green curve) and 300°C (the blue curve).

Since $T_C$ is significantly greater than $T_A$, $x_C < x_A$ and $0 < \eta_0 \ll 1$ (obtained by the fitting of the experiment [2]), the absolute value $|\eta(T,x)|$ is significantly smaller than the absolute value $|\alpha(T,x)|$ for $T << T_C$ and $x < x_C$ (e.g., $\alpha(T,x) \leq -1$). The smallness of $\eta(T,x)$ is responsible for the smallness of $A_S$ in the AFE and FEI phase. In particular, $A_S \ll P_S$ almost in the entire region of the FEI phase, where $|\alpha(T,x)| \gg |\eta(T,x)|$. Notably that the small positive values of $\xi$ and small $A_S$ (in comparison with $P_S$) define the very weak influence of the antipolar order on the polar order in the FEI phase. In result, $P_S$ is almost insensitive to the appearance of $A_S$ in the FEI phase and its increase near the FEI-AFE boundary (compare the color maps in **Fig. 1(b)** and **1(c)**). Since $P_S$ is not small in almost the entire region of the FE and FEI phases, the converse statement is not true, namely the influence of $P_S$ on $A_S$ can be rather strong when $\alpha(T,x)$ is negative and $|\alpha(T,x)| \gg |\eta(T,x)|$. In result the spontaneous polarization suppresses the antipolar ordering in the FE phase. Since the influence of $\vec{A}$ on $\vec{P}$ appeared negligibly small, the linear dielectric susceptibility $\chi$ calculated at $E \to 0$ is also insensitive to the appearance and changes of $\vec{A}$ (see **Fig. 1(d)**). Note that hereinafter we consider the case of collinear $\vec{E}$ and $\vec{P}$ vectors, which allows us to omit the subscripts in the susceptibility tensor component, e.g., put $\chi \equiv \chi_{33}$.

Because the sixth order terms, e.g., $\frac{\gamma}{4}(P^6 + A^6)$, are omitted in Eq.(1), the phase diagram in **Fig. 1(a)** does not contain the morphotropic phase boundary (MPB) observed experimentally between FE (or FEI) and NP phase and presented in fig.1(a) in Ref.[2], as well as more complex types of mixed phases observed experimentally in thin strained films of $Bi_{1-x}Sm_xFeO_3$ [18]. Moreover, the oversimplified 2-4-power Landau expansion (1), which does not contain the gradient and flexoelectric energies, cannot describe the possible appearance of the spatially-modulated incommensurate (IC) phases inherent to the rare-earth doped $BiFeO_3$ [2, 17]. To use the simple analytical expressions for the energies of different phases listed in **Table A1** [32], hereinafter we use the oversimplified free energy (1) instead of more rigorous 2-4-6-power LGD functionals with the gradient and flexoelectric terms, which are listed in Refs. [2, 17, 18]. The absence of the sixth order terms in the Landau energy (1) allows the analytical description of the phase diagrams of $Bi_{1-x}Sm_xFeO_3$ nanoparticles, which is presented below.

## B. Analytical calculations of the phase diagrams and polar properties of $Bi_{1-x}Sm_xFeO_3$ nanoparticles

To determine the phase diagrams of the stress-free $Bi_{1-x}Sm_xFeO_3$ nanoparticles we use the FSM-LGDK-SH approach. Corresponding free energy functional $F$ includes a bulk part, $F_{bulk}$, a polarization gradient energy contribution, $F_{grad}$, elastic and flexoelectric energy, $F_{el+fl}$, an electrostatic contribution, $F_{el}$, and a surface energy, $F_S$:

$$F = F_{bulk} + F_{grad} + F_{el+fl} + F_{el} + F_S, \tag{3}$$

where the constituent parts are

$$F_{bulk} = \int_V dV \left\{\frac{1}{2}\alpha P^2 + \frac{1}{2}\eta A^2 + \frac{1}{4}\beta_{ij}\left(P_i^2 P_j^2 + A_i^2 A_j^2\right) + \frac{1}{2}\xi_{ij} P_i^2 A_j^2\right\}, \tag{4a}$$

$$F_{grad} = \int_V dV \frac{g_{ijkl}}{2}\left(\frac{\partial P_i}{\partial x_j}\frac{\partial P_k}{\partial x_l} + \frac{\partial A_i}{\partial x_j}\frac{\partial A_k}{\partial x_l}\right), \tag{4b}$$

$$F_{el+fl} = \int_V dV \left\{\frac{F_{ijkl}}{2}\left(\frac{\partial P_i}{\partial x_j}\sigma_{kl} - \frac{\partial \sigma_{kl}}{\partial x_j}P_i\right) - \frac{s_{ijkl}}{2}\sigma_{ij}\sigma_{kl}\right\}, \tag{4c}$$

$$F_{el} = -\int_V dV \left(P_i E_i + \frac{\varepsilon_0 \varepsilon_b}{2}E^2\right), \tag{4d}$$

$$F_S = \frac{1}{2}\int_S dS \ \left(c_P P^2 + c_A A^2\right). \tag{4e}$$

Here $V$ is the volume and $S$ is the surface of the $Bi_{1-x}Sm_xFeO_3$ nanoparticle. Symbols $dV$ and $dS$ designate the volume and surface differentials for the volume and surface integration, respectively. Polar and antipolar order parameter vectors are $\vec{P} = (P_1, P_2, P_3)$ and $\vec{A} = (A_1, A_2, A_3)$, and their magnitudes are $P^2 = P_1^2 + P_2^2 + P_3^2$ and $A^2 = A_1^2 + A_2^2 + A_3^2$. The LGDK expansion coefficients $\alpha$ and $\eta$ are given by Eq.(2); $g_{ijkl}$ is the polarization gradient tensor, $F_{ijkl}$ is the flexoelectric tensor, $\sigma_{kl}$ are elastic stresses, $s_{ijkl}$ are the components of the elastic compliance tensor, $c_P$ and $c_A$ are the surface energy coefficients; $\varepsilon_0$ is a universal dielectric constant, $\varepsilon_b$ is the relative background permittivity [33]; $E_i$ are the electric field components. Hereinafter we use the Einstein summation rule over the repeated subscripts $i$ and/or $j$, where $i, j$ =1, 2 and 3.

The absence of the sixth order terms in the Landau energy (4a), flexoelectric and elastic energies (4c) in the assumption of negligibly small stresses $\sigma_{kl}$ and their gradients $\frac{\partial \sigma_{kl}}{\partial x_j}$, allows the analytical description of the phase diagrams of $Bi_{1-x}Sm_xFeO_3$ nanoparticles without the MPB and IC phases, which are observed in bulk samples and thin films of $Bi_{1-x}Sm_xFeO_3$ [2, 17, 18]. The MPB and IC phases in the $Bi_{1-x}Sm_xFeO_3$ nanoparticles can be calculated numerically using e.g., the finite element modelling (FEM) or the phase-field approach [34] with mandatory inclusion of the flexoelectric and elastic energies [35]. The complex and interesting problem is beyond the scope of this work.

The polar order parameter $P_i$ is readily measurable in the electric field $E_i$, but not the antipolar order parameter $A_i$. The influence of $A_i$ on the field dependence of $P_i$ can be observed only for the high strength of the biquadratic coupling $\xi_{ij}$ between $P_i$ and $A_i$ (see e.g., **Fig. A1** and **A2** in **Appendix A** [32]). For small $\xi_{ij}$ the term $\frac{1}{2}\xi_{ij}P_i^2A_j^2$ in Eq.(4a) cannot change the dielectric susceptibility, $\chi_{ij}=\frac{\partial P_i}{\partial E_{ij}}$, in a noticeable way.

The quasi-static electric field $E_i$ is related to the electric potential $\varphi$ as $E_i=-\frac{\partial\varphi}{\partial x_i}$. The potential $\varphi$ satisfies the Poisson equation inside the $Bi_{1-x}Sm_xFeO_3$ nanoparticle and obeys the Debye-Huckell equation inside the screening shell with free charge carriers (mobile ions, electrons and/or vacancies) covering the nanoparticle (see **Fig. 2(a)**).

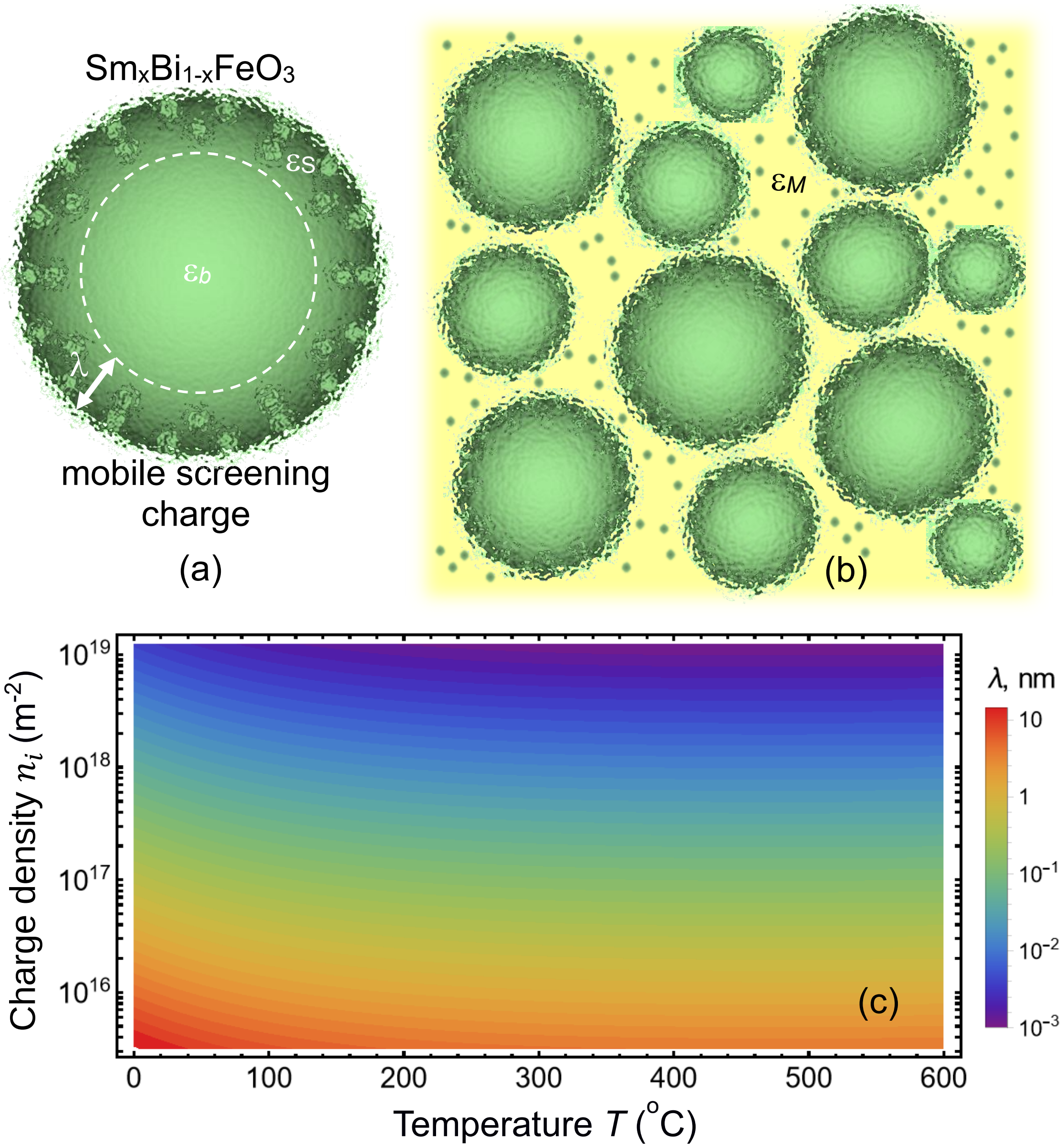

**FIGURE 2**. **(a)** The radial cross-section of the spherical $Bi_{1-x}Sm_xFeO_3$ nanoparticle covered with the shell of ionic-electronic screening charges with the effective screening length $\lambda$**. (b)** $Bi_{1-x}Sm_xFeO_3$ nanoparticles in the environment with free ionic-electronic charges. **(c)** The dependence of λ on the temperature $T$ and the surface charge density $n_i$ calculated for $\Delta G_1 = \Delta G_2 \cong 0.1$ eV and $Z_1 = -Z_2 = 1$.

Spatial-temporal evolution of the polar and antipolar order parameters, $P_i$ and $A_i$, is determined from the coupled time-dependent Euler-Lagrange equations, obtained by the minimization of the free energy $F$. These equations have a relatively simple form:

$$\Gamma_P \frac{\partial P_i}{\partial t} + \alpha P_i + \beta_{ij} P_i P_j^2 + \xi_{ij} A_i^2 P_i - g_{ijkl} \frac{\partial P_k}{\partial x_j \partial x_l} = E_i, \tag{5a}$$

$$\Gamma_A \frac{\partial A_i}{\partial t} + \eta A_i + \beta_{ij} A_i A_j^2 + \xi_{ij} P_i^2 A_i - g_{ijkl} \frac{\partial A_k}{\partial x_j \partial x_l} = 0. \tag{5b}$$

Here the summation over "$i$" is absent, $\Gamma_P$ and $\Gamma_A$ are Landau-Khalatnikov relaxation coefficients [36]. The relaxation times of $P_i$ and $A_i$ are $\tau_P = \Gamma_P/|a_P|$ and $\tau_A = \Gamma_A/|a_A|$, respectively.

Corresponding boundary conditions to Eqs.(5) are of the third kind [37]:

$$\left.\left(c_P P_i - g_{ijkl} e_j \frac{\partial P_k}{\partial x_l}\right)\right|_S = 0, \quad \left.\left(c_A A_i - g_{ijkl} e_j \frac{\partial A_k}{\partial x_l}\right)\right|_S = 0. \tag{6}$$

Here $\vec{e}$ is the outer normal to the nanoparticle surface $S$. Sometimes it is convenient to introduce the extrapolation lengths, $\Lambda_P = \frac{g}{c_P}$ and $\Lambda_A = \frac{g}{c_A}$, which physical range is (0.5 – 5) nm [38].

For the sake of simplicity let us consider a single-domain spherical $Bi_{1-x}Sm_xFeO_3$ nanoparticle, in which the direction of the spontaneous polarization $\overrightarrow{P_S}$ is colinear with the axis 3. For the geometry the electric field component $E_3$ is a superposition of external and depolarization fields, $E_3^e$ and $E_3^d$, respectively. For the single-domain nanoparticle the analytical expressions for the electric field components have the form:

$$E_3^d = -\frac{1}{\varepsilon_b + 2\varepsilon_M + \varepsilon_s (R/\lambda)} \frac{P_3}{\varepsilon_0}, \qquad E_3^e = \frac{3\varepsilon_M}{\varepsilon_b + 2\varepsilon_M + \varepsilon_s (R/\lambda)} E_3^0, \tag{7}$$

where $R$ is the radius of the nanoparticle, λ is the "effective" screening length in the conductive shell with the relative dielectric permittivity $\varepsilon_s$; $\varepsilon_M$ is the relative effective dielectric permittivity of the environment (see **Fig. 2(b)**). The derivation of Eqs.(7) is given in Ref. [39].

If $\lambda \gg R$ and $\varepsilon_M \sim \varepsilon_b$, the field inside the nanoparticle is of the same order as the applied field $E_3^0$. However, λ can be rather small (less than 0.1 – 1 nm) due to the high density of the free surface charges in the conducting shell [40], which, in turn, appear due to the strong band-bending caused by the "bare" depolarization field [41].

Within the SH model the dependence of the surface charge density $\sigma_S[\phi]$ on electric potential excess $\delta\phi$ at the surface of the nanoparticle is controlled by the concentration of positive and negative surface charges $\theta_i[\phi]$ in a self-consistent manner. The density $\sigma_S[\phi]$ obeys the Langmuir adsorption isotherm [42, 43]:

$$\sigma_S[\phi] = \sum_i \frac{eZ_i\theta_i[\phi]}{N_i} \cong \sum_i \frac{eZ_i}{N_i}\left(1+\exp\left[\frac{\Delta G_i + eZ_i\delta\phi}{k_B T}\right]\right)^{-1}, \qquad (8)$$

where $e$ is the electron charge, $Z_i$ is the ionization number of the surface ions/vacancies, $T$ is the absolute temperature, $1/N_i$ are the densities of positive and negative charge species in saturation ($i$=1,2), $\Delta G_i$ are the formation energies of the surface charges (e.g., ions and/or vacancies) at normal conditions. It is reasonable to assume that $\varepsilon_M \approx \varepsilon_s$ for ultra-thin shells consisting of ionic-electronic charge adsorbed from the environment.

The linearization of expression (8) for small built-in potentials, $\left|\frac{eZ_i\delta\phi}{k_B T}\right| < 1$, leads to the approximate expression for the effective surface charge density $\sigma_S$ and inverse screening length λ [43]:

$$\sigma_S[\phi] \approx -\varepsilon_0 \frac{\delta\phi}{\lambda}, \qquad \frac{1}{\lambda} \approx \sum_i \frac{(eZ_i)^2 n_i}{4\varepsilon_0 k_B T \cosh^2\left(\frac{\Delta G_i}{2k_B T}\right)}. \qquad (9)$$

Here $n_i = 1/N_i$ is the surface charge density. The dependence of λ on the temperature $T$ and the surface charge density $n_i$ is shown in **Fig. 2(c)**. Hereinafter we use typical and equal values of the surface charge formation energies, $\Delta G_1 = \Delta G_2 \cong$0.1 eV, opposite ionization numbers, $Z_1 = -Z_2$ =1 and vary $n_i$ from $10^{16}$m$^{-2}$ to $10^{19}$m$^{-2}$. The chosen range of the surface charge parameters are consistent with the literature data [20, 21].

When λ is very small (e.g., less than 0.1 nm), the shell provides very good screening of the $Bi_{1-x}Sm_xFeO_3$ spontaneous polarization and thus prevents the domain formation. Thus, the assumption of the single-domain state in the $Bi_{1-x}Sm_xFeO_3$ nanoparticle is self-consistent for e.g., $n_i \geq 2\cdot 10^{17}$m$^{-2}$. For λ ≥0.1 nm one should use the FEM to account for the possible domain formation [34, 44].

Substitution of Eqs.(7) in Eqs.(5) leads to the following renormalization of the dimensionless coefficient $\alpha(T,x)$ in Eq.(4a):

$$\alpha_R(T,x,R) = \alpha(T,x) + \frac{(\alpha_0\varepsilon_0)^{-1}}{\varepsilon_b + 2\varepsilon_M + \varepsilon_S(R/\lambda)}. \qquad (10a)$$

Next, substitution of Eq.(2a) and Eq.(9) in Eqs.(10a) leads to the approximate expression for the renormalized coefficient $\alpha_R(T,x,R)$:

$$\alpha_R(T,x,R) \approx \frac{T}{T_C} - \exp\left[-\left(\frac{x}{x_C}\right)^4\right] + \frac{(\alpha_0\varepsilon_0)^{-1}}{\varepsilon_b + 2\varepsilon_M + \sum_i \frac{\varepsilon_S R (eZ_i)^2 n_i}{4\varepsilon_0 k_B T \cosh^2\left(\frac{\Delta G_i}{2k_B T}\right)}}. \qquad (10b)$$

The free energy (3) with the renormalized coefficient $\alpha_R(T,x,R)$ allows analytical description of the size effects and ferro-ionic coupling influence on the phase diagrams and polar properties of the single-domain $Bi_{1-x}Sm_xFeO_3$ nanoparticles. Corresponding analytical expressions for the energies of different phases, spontaneous polarization $P_S$ and antipolar long-range order $A_S$ are listed in **Table A1** [32], where the renormalized coefficient $\alpha_R(T,x,R)$ is used. The influence of finite size effect and ferro-ionic coupling is determined by the ratio $R/\lambda$, and the role of the environment is determined by its permittivity $\varepsilon_M$ in the denominator of Eq.(10).

The dimensionless dielectric susceptibility is given by expression:

$$\chi = \frac{\partial P_3}{\partial E_3} = \frac{\eta + 3\beta A_3^2 + \xi P_3^2}{(\alpha_R + 3P_3^2 + \xi A_3^2)(\eta + 3\beta A_3^2 + \xi P_3^2) - 4\xi^2 P_3^2 A_3^2}. \quad (11)$$

Hereinafter we re-designate $\chi_{33} \to \chi$ for the sake of brevity.

**B.1. The influence of the size effect on phase diagrams and polar properties of $Bi_{1-x}Sm_xFeO_3$ nanoparticles**

Phase diagrams of the spherical $Bi_{1-x}Sm_xFeO_3$ nanoparticles, calculated in dependence of the Sm content $x$ and temperature $T$ for different values of the particle radius $R$ are shown in **Fig. 3(a)-(d)**. The diagrams are calculated for typical parameters of the surface charge, relative background permittivity $\varepsilon_b$ =10 and high-k environment with relative dielectric permittivity $\varepsilon_M$ =30. The diagram calculated for a relatively big nanoparticle (**Fig. 3(a)**) looks like the diagram of a bulk material (**Fig. 1(a)**) except for the small region of the reentrant FE phase appearing at small $x < 0.025$. The reentrant phase is induced by the strong enough ferro-ionic coupling in the nanoparticle. A relatively large region of the FE phase exists for $0 \leq x \leq 0.1$, but its area decreases significantly with $R$ decrease from 50 nm to 2.5 nm (compare **Fig. 3(a), 3(b), 3(c)** and **3(d)**, respectively). The two regions of the FEI phase are relatively small and exists for $0 \leq x \leq 0.1$, but they merge and increase significantly with $R$ decrease from 50 nm to 2.5 nm (compare **Fig. 3(a), 3(b), 3(c)** and **3(d)**, respectively). In other words, the reentrant FEI phase partially substitutes the FE phase with $R$ decrease. Also, the area of the AFE phase region ($0.1 \leq x \leq 0.15$) increases slightly with $R$ decrease, at that the AFE phase partially substitute the FEI phase. For $x > 0.15$ the NP phase is stable.

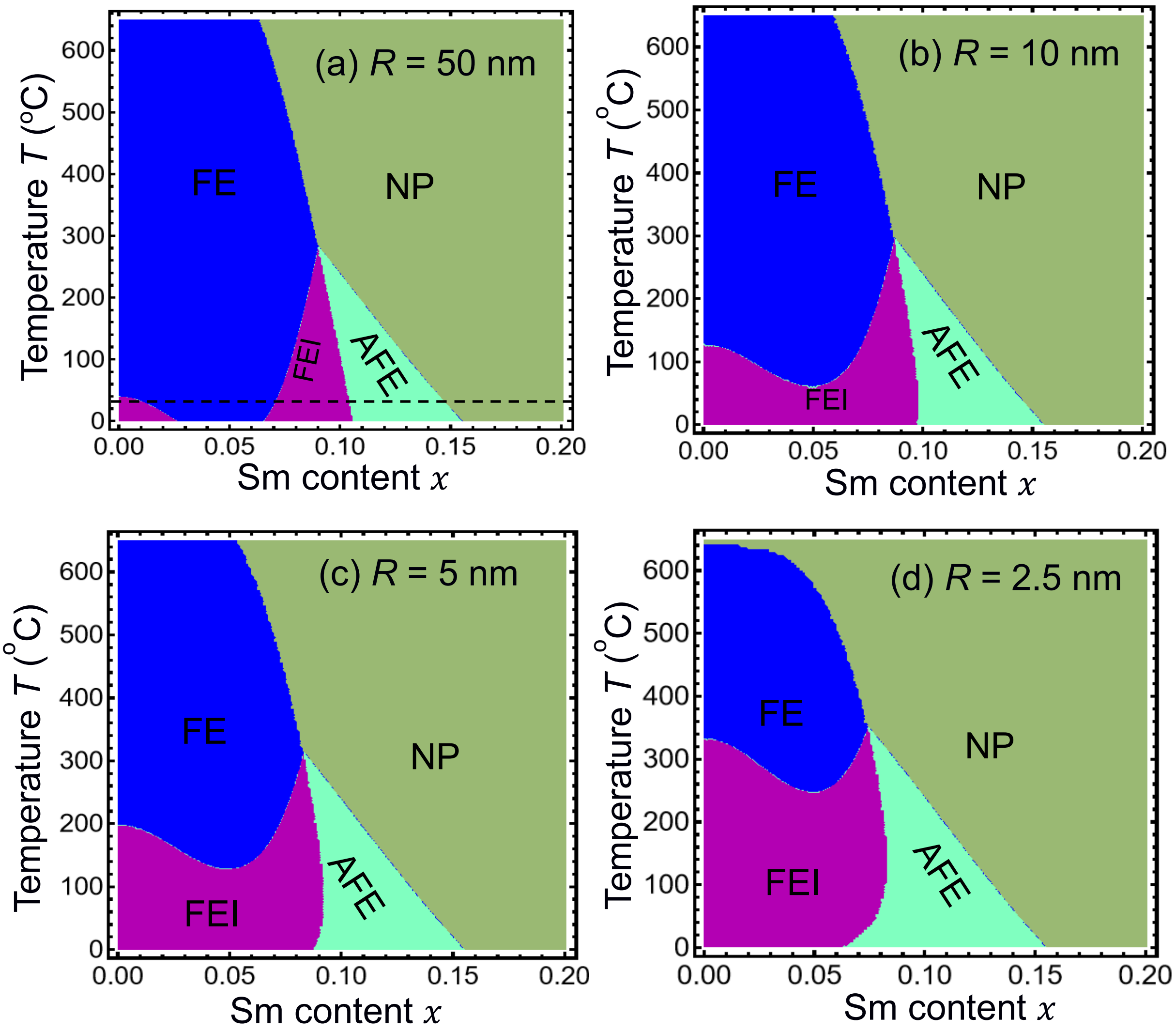

**FIGURE 3.** Phase diagrams of $Bi_{1-x}Sm_xFeO_3$ nanoparticles in dependence on the Sm content $x$ and temperature $T$. The diagrams are calculated using the free energy (3) for the following parameters: $\alpha_0 = 1.13 \cdot 10^8$ m/F, $T_C = 1100$ K, $T_A = 800$ K, $x_C = 0.1$, $x_A = 0.15$, $\eta_0 = 0.138$, $\beta = 1$, and $\xi = 0.121$, $\Delta G_i$ =0.1 eV, $Z_1 = -Z_2$ =1, $n_i = 10^{18}$m$^{-2}$, $\varepsilon_b$ =10, $\varepsilon_M$ =30, and several values of the particle radius $R$ = 50 nm **(a)**, 10 nm **(b)**, 5 nm **(c)** and 2.5 nm **(d)**. The dashed horizontal line in the plot (a) corresponds to 300 K.

The reentrant transition between the FEI and FE phases exists for $R > R_R$ as shown in **Fig. 4(a)** (e.g., $R_R \approx 13$ nm at $T$ =25°C, $\varepsilon_M$ =30 and $n_i = 10^{18}$m$^{-2}$). The influence of $R$ on the transition from the FEI to the AFE phase is much weaker; and the influence of $R$ on the AFE-NP transition is absent, because the depolarization field is absent in the homogeneous AFE and NP phases.

Dependencies of the spontaneous polarization $P_S$ and antipolar parameter $A_S$ on the Sm content $x$ and particle radius $R$ are shown in **Fig. 4(b)** and **Fig. 4(c)**, respectively. Since the biquadratic coupling strength $\xi$ in Eq.(4a) is positive and very small (from the best fitting of a bulk phase diagram [2]), the spontaneous polarization does not have any visible features at the boundary between the FE and the FEI phases. It decreases significantly approaching the AFE boundary and disappears in the AFE phase. At that the boundary between the FEI and AFE phases corresponds to the second order phase transition. The spontaneous polarization relatively weakly depends on $R$ for $R > 25$ nm. A noticeable decrease of $P_S$ exists at $R < 5$ nm only. Here it is maximal at $x = 0$, decreases monotonically with $x$ increase and eventually disappears at $x \geq x_{cr}(R)$, where $x_{cr}(R) \approx 0.1$ for $R > 5$ nm. The spontaneous antipolar order exists in the FEI and AFE phases. It gradually disappears approaching the FEI-FE boundary (which has a parabolic-like shape), as well as approaching the AFE-NP boundary (which is close to the vertical line $x \approx 0.15$). The antipolar order reaches maximal values at the FEI-AFE boundary.

The dependences of the nanoparticle static dielectric susceptibility $\chi$ on the Sm content $x$ calculated in zero electric field ($E \rightarrow 0$) for several values of the nanoparticle radius from 2.5 nm to 50 nm are shown in **Fig. 4(d)**. The susceptibility diverges at the critical value $x = x_{cr}$, and $x_{cr}$ increases with increase in $R$ (compare different curves in **Fig. 4(d)**). The divergency corresponds to the second order transition between the FEI and AFE phases. Blue arrows in **Fig. 4(d)** point on the tiny fractures, which correspond to the reentrance of the FEI phase for the 50-nm $Sm_xBi_{1-x}FeO_3$ nanoparticle. For $R < R_R$ any features of the reentrant FE-FEI transition are not visible at the dielectric susceptibility curve, because the transition is absent for the radii range in **Fig. 4(a).** Since the strength $\xi$ of the biquadratic coupling term in Eq.(4a) is very small, the appearance and changes of the antipolar order cannot change and the dielectric susceptibility $\chi$ in a noticeable way.

To summarize, the size effect of the spontaneous polarization becomes significant for the small radius (less than 10 nm) near the FEI-AFE boundary. The size effect of the antipolar long-range order is significant near the “bottom” of the reentrant FEI-FE boundary (for radii 15 – 20 nm).

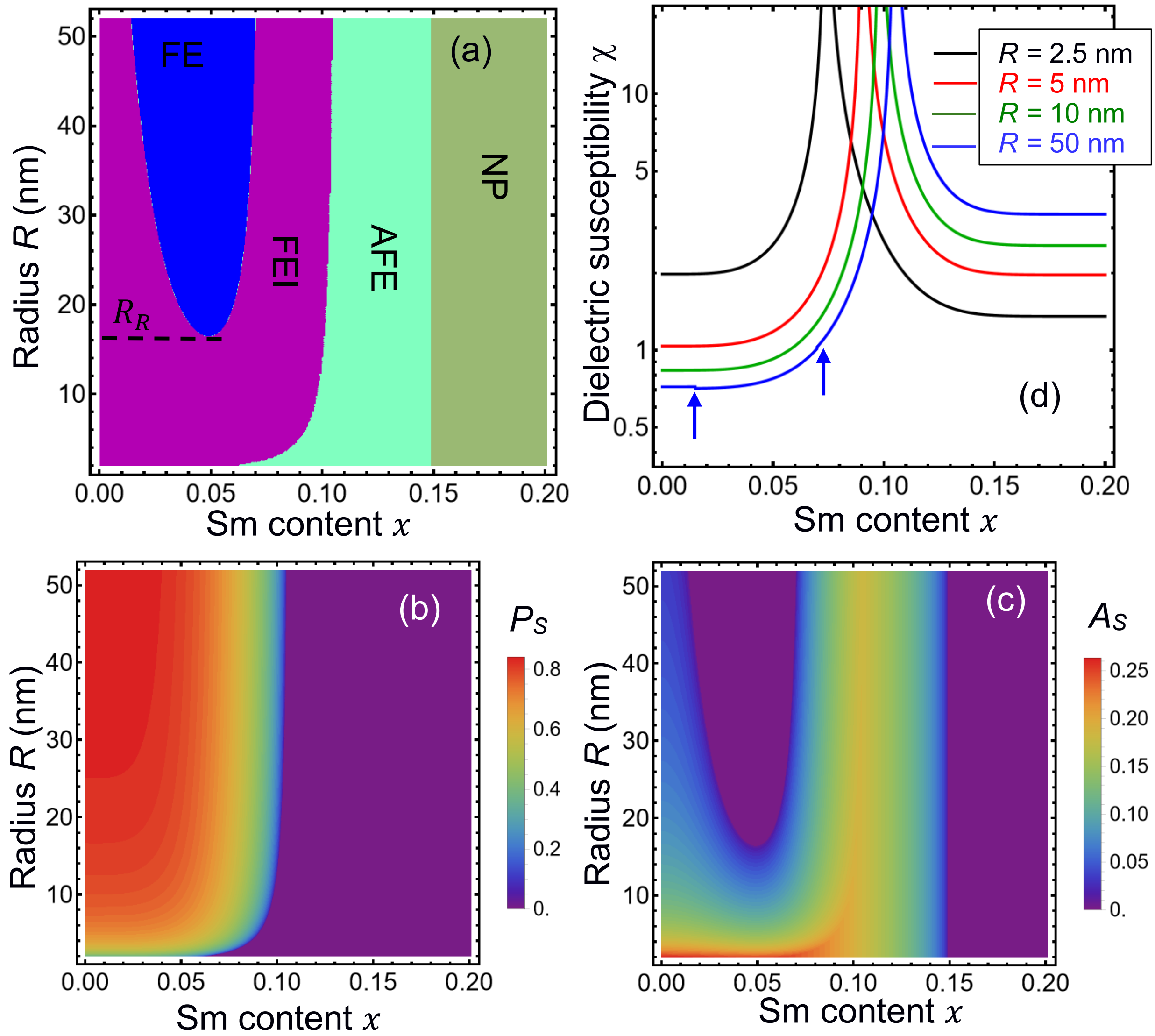

**FIGURE 4. (a)** The phase diagram of $Bi_{1-x}Sm_xFeO_3$ nanoparticles in dependence on the Sm content $x$ and particle radius $R$. Dependencies of the spontaneous polarization $P_S$ **(b)** and antipolar order $A_S$ **(c)** on the x and $R$. **(d)** The dependence of the nanoparticle dielectric susceptibility on the Sm content $x$ calculated at $E = 0$ for $R$ =2.5 nm (the black curve), 5 nm (the red curve), 10 nm (the green curve) and 50 nm (the blue curve). Blue arrows point on the tiny fractures, which correspond to the reentrance of the FEI phase. The temperature $T$ =25°C, $\varepsilon_M$ =30; other parameters are the same as in **Fig. 3.**

## B.2. The influence of the environment dielectric permittivity on phase diagrams and polar properties of $Bi_{1-x}Sm_xFeO_3$ nanoparticles

Phase diagrams of the small $Bi_{1-x}Sm_xFeO_3$ nanoparticles, calculated in dependence of the Sm content $x$ and temperature $T$ for different values of the environment relative dielectric permittivity $\varepsilon_M$ are shown in **Fig. 5.** The value $\varepsilon_M = 300$ corresponds to the paraelectric matrix (e.g., $SrTiO_3$), $\varepsilon_M = 81$ corresponds to the water-based electrolyte, $\varepsilon_s = 10$ is for the case when the nanoparticle and the environment have the same dielectric constants, and $\varepsilon_s = 3$ corresponds to the low-k environment. The diagram of the 5-nm $Bi_{1-x}Sm_xFeO_3$ nanoparticle calculated for $\varepsilon_M = 300$ is very similar to the diagram of a bulk $Bi_{1-x}Sm_xFeO_3$ (compare **Fig. 5(a)** and **Fig. 1(a)**), except for the small region of the reentrant FEI phase appearing in the nanoparticle at small $x < 0.025$. The similarity with the bulk $Bi_{1-x}Sm_xFeO_3$ comes from the small contribution of the depolarization field for high $\varepsilon_M$, because the field is proportional to $1/\varepsilon_M$ in accordance with Eq.(7). The relatively large region of the FE phase, which exists for $0 \leq x \leq 0.1$ and $\varepsilon_M \geq 80$, decreases significantly with $\varepsilon_M$ decrease. In addition to the decrease of the FE phase region, the decrease in $\varepsilon_M$ from 300 to 3 leads to the gradual substitution of the FE phase by the FEI phase and then to the complete disappearance of the FEI phase, which transforms into the AFE phase (compare **Fig. 5(a), 5(b), 5(c)** and **5(d)**, respectively).

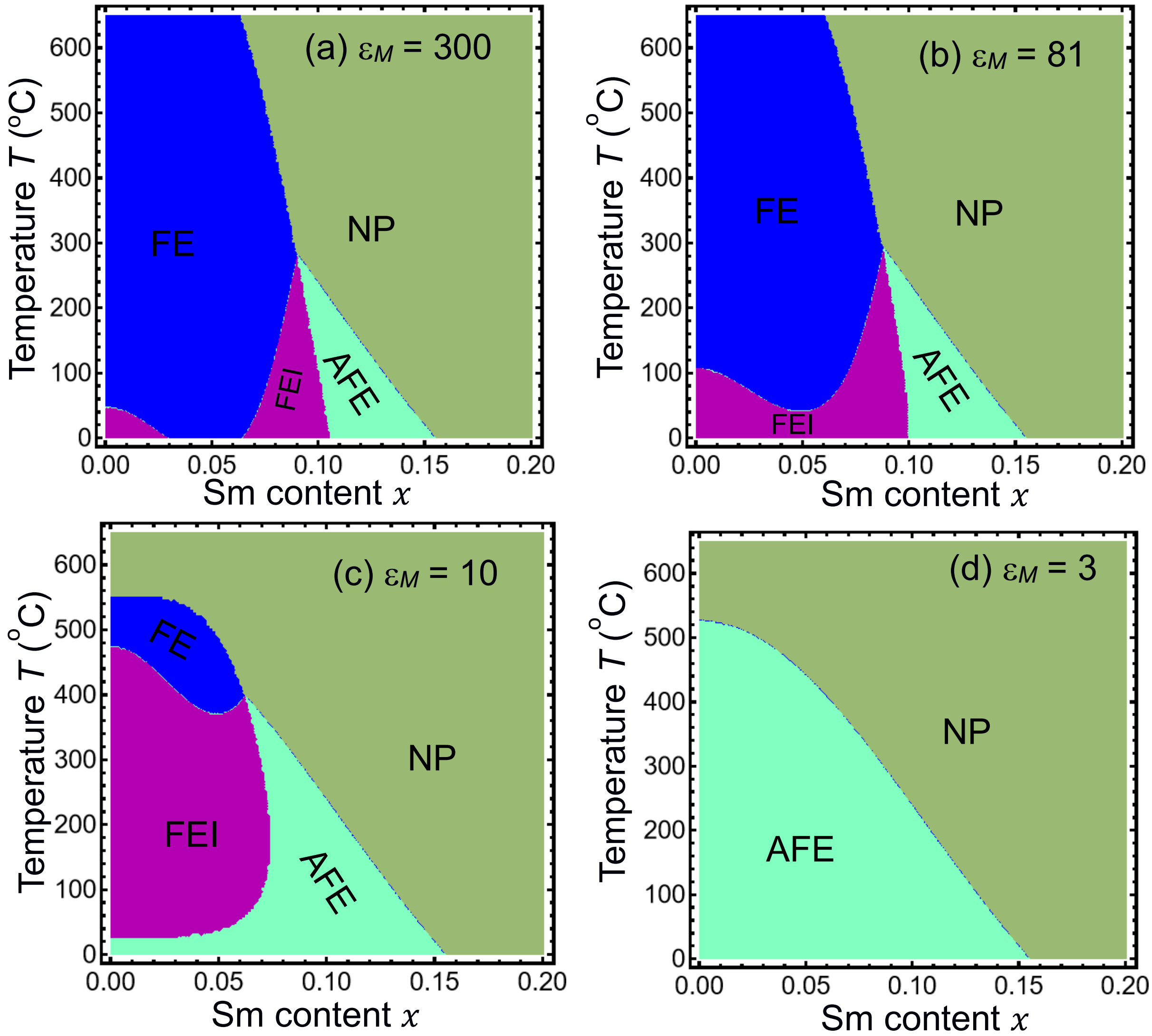

**FIGURE 5.** Phase diagrams of the $Bi_{1-x}Sm_xFeO_3$ nanoparticles in dependence on the Sm content $x$ and temperature $T$. The diagrams are calculated using the free energy (3) for $R = 5$ nm, $n_i = 10^{18}\text{m}^{-2}$, and several values of the environment relative dielectric permittivity $\varepsilon_M$ =300 **(a)**, 81 **(b)**, 10 **(c)** and 3 **(d)**. Other parameters are the same as in **Fig. 3**.

The reentrant transition between the FEI and FE phases appears for $\varepsilon_M$ bigger than the characteristic value $\varepsilon_R$, as shown in **Fig. 6(a)** (e.g., $\varepsilon_R \approx 100$ at $T$ =25°C, $R = 5$ nm and $n_i = 10^{18}$m$^{-2}$). The influence of $\varepsilon_M$ on the FEI-AFE transition is almost absent for $\varepsilon_M > 50$, and the influence of $\varepsilon_M$ on the AFE-NP transition is absent, because the depolarization field is absent in the homogeneous AFE and NP phases.

Dependencies of the spontaneous polarization $P_S$ and antipolar order $A_S$ on the Sm content $x$ and relative dielectric permittivity $\varepsilon_M$ are shown in **Fig. 6(b)** and **Fig. 6(c)**, respectively. Similarly to the situation shown in **Fig. 4(b)**, the spontaneous polarization does not have any features at the FEI-FE phase boundary due to the small value of the biquadratic coupling strength $\xi$. $P_S$ decreases significantly approaching the AFE boundary and disappears in the AFE phase. Inside the regions of the FE and FEI phases the spontaneous polarization relatively weakly depends on $\varepsilon_M$ for $\varepsilon_M > 50$. At $\varepsilon_M < \varepsilon_{cr}$ the transition from the FEI to the AFE phase occurs due to the high depolarization effects. The critical value $\varepsilon_{cr} \approx 8$ for the chosen parameters $T$ =25°C, $R = 5$ nm and $n_i = 10^{18}$m$^{-2}$. $P_S$ is maximal for $x = 0$, decreases monotonically with $x$ increase and eventually disappears for $x \geq x_{cr}(\varepsilon_M)$. At that $x_{cr}$ becomes very close to the "bulk" value $x_C = 0.10$ for big $\varepsilon_M$.

Similarly to the situation shown in **Fig. 4(c)**, the spontaneous antipolar order parameter $A_S$ exists in the FEI and AFE phases. It gradually disappears approaching the FEI-FE boundary (which has a parabolic-like shape), as well as approaching the AFE-NP boundary (which is close to the vertical line $x \approx x_A = 0.15$). The parameter $A_S$ has a diffuse local maximum at the FEI-AFE boundary and reaches the highest values in the AFE phase for the small $\varepsilon_M \leq \varepsilon_{cr}$ and $x < 0.05$. Exactly the region of maximal $A_S$ is the region where $P_S$ is absent due to the high depolarization field. Hence this figure confirms the significant influence of the polar order on the antipolar order, and the negligibly small influence of $A_S$ on $P_S$ due to the small positive $\xi$ and $\eta$ values.

The dependences of the nanoparticle dielectric susceptibility on the Sm content $x$ calculated at $E = 0$ for several values of the dielectric permittivity $\varepsilon_M$ are shown in **Fig. 6(d)**. For the very small $\varepsilon_M < \varepsilon_{cr}$ the susceptibility monotonically decreases with increase in $x$, because the FEI-AFE transition is absent. For $\varepsilon_M \geq \varepsilon_{cr}$ the susceptibility diverges at the critical value $x = x_{cr}$, and $x_{cr}$ increases with increase in $\varepsilon_M$ (compare different curves in **Fig. 6(d)**). The divergency corresponds to the second order transition between the FEI and AFE phases. Blue arrows in **Fig. 6(d)** point on the tiny fractures, which correspond to the reentrance of the FEI phase for the $\varepsilon_M = 300$. For $\varepsilon_M < \varepsilon_R$ any features of the reentrant FE-FEI transition are not visible at the dielectric permittivity curve, because the transition is absent for $\varepsilon_M < \varepsilon_R$ in **Fig. 6(a)**.

The influence of the environment dielectric permittivity $\varepsilon_M$ on the spontaneous polarization becomes significant for $\varepsilon_M < 50$ near the FEI-AFE boundary. The influence of $\varepsilon_M$ on the antipolar order is the most significant near the "bottom" of the reentrant FEI-FE boundary (e.g., for $\varepsilon_M \approx 50 - 100$ at $R$ =5 nm and $n_i = 10^{18}$m$^{-2}$).

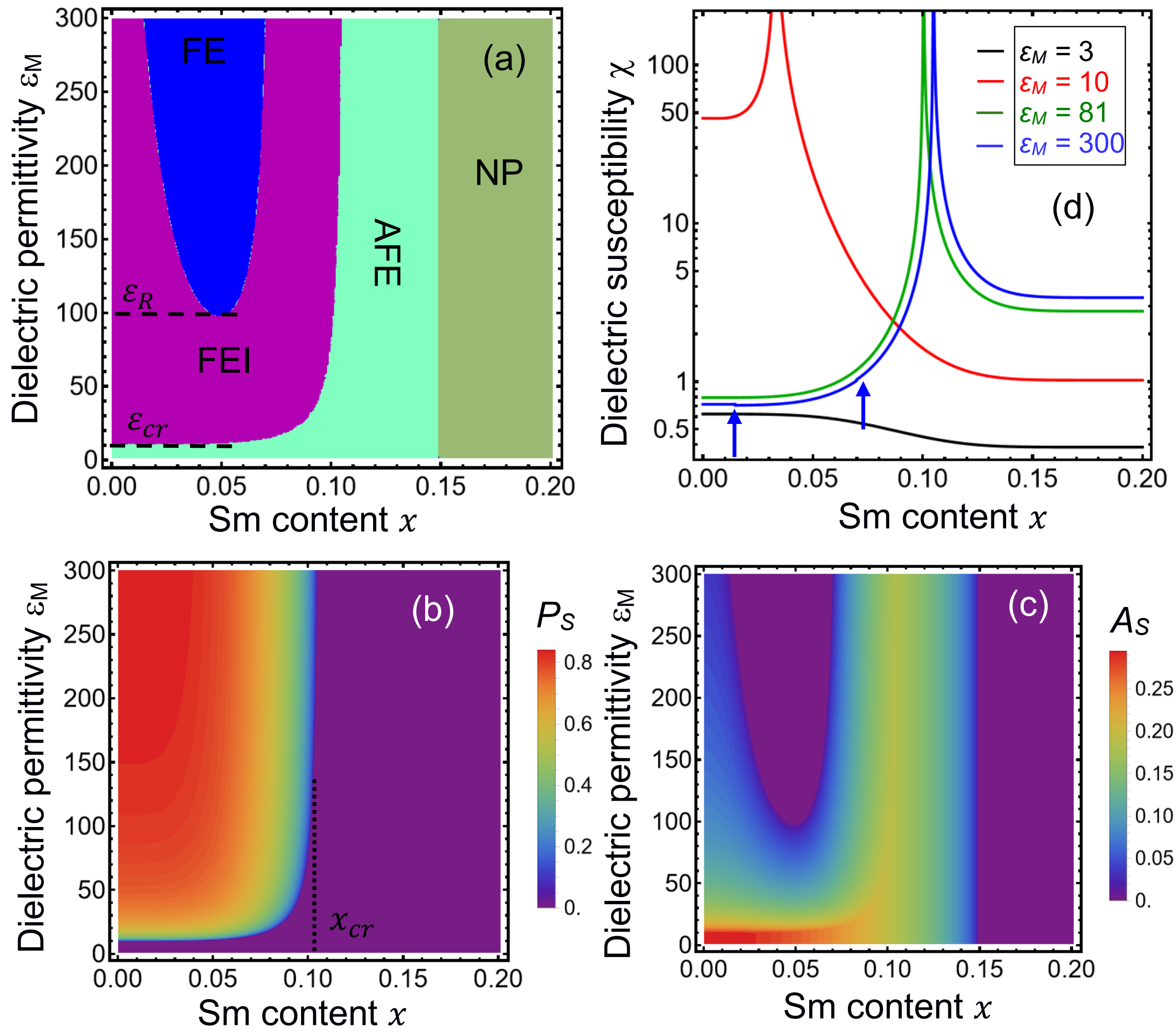

**FIGURE 6. (a)** The phase diagram of $Bi_{1-x}Sm_xFeO_3$ nanoparticles in dependence on the Sm content $x$ and relative dielectric permittivity $\varepsilon_M$ of environment. Dependencies of the spontaneous polarization $P_S$ **(b)** and antipolar order $A_S$ **(c)** on $x$ and $\varepsilon_M$. **(d)** The dependence of the nanoparticle dielectric susceptibility on the Sm content $x$ calculated at $E = 0$ for $\varepsilon_M = 3$ (the black curve), 10 (the red curve), 81 (the green curve) and 300 (the blue curve). Blue arrows point on the tiny fractures, which correspond to the reentrance of the FEI phase. The temperature $T$ =25°C, $R = 5$ nm and $n_i = 10^{18}$m$^{-2}$; other parameters are the same as in **Fig. 3.**

### B.3. The influence of the ferro-ionic coupling on phase diagrams and polar properties of $Bi_{1-x}Sm_xFeO_3$ nanoparticles

Phase diagrams of the small $Bi_{1-x}Sm_xFeO_3$ nanoparticles, calculated in dependence of the Sm content $x$ and temperature $T$ for different values of the surface ionic-electronic charge density $n_i$, $R = 5$ nm and $\varepsilon_M = 10$, are shown in **Fig. 7**. The diagram of the 5-nm nanoparticle calculated for the relatively high $n_i = 10^{19}\text{m}^{-2}$ is close to the diagram of a bulk material (compare **Fig. 7(a)** and **Fig. 1(a)**), except for the curved region of the reentrant FEI phase appearing at small $x < 0.1$. The similarity with a bulk is conditioned by the strong ferro-ionic coupling, which provides the almost perfect screening of the depolarization field at high $n_i$, since the screening length λ is proportional to $1/n_i$ in accordance with Eq.(8). The relatively large region of the FE phase, which exists for high $n_i$, decreases significantly with $n_i$ decrease. In particular, the decrease in $n_i$ from $10^{19}\text{m}^{-2}$ to $5 \cdot 10^{17}\text{m}^{-2}$ leads to the gradual substitution of the FE phase by the FEI phase and then to the gradual disappearance of the FEI phase, which transforms into the AFE phase (compare **Fig. 7(a), 7(b), 7(c)** and **7(d)**, respectively).

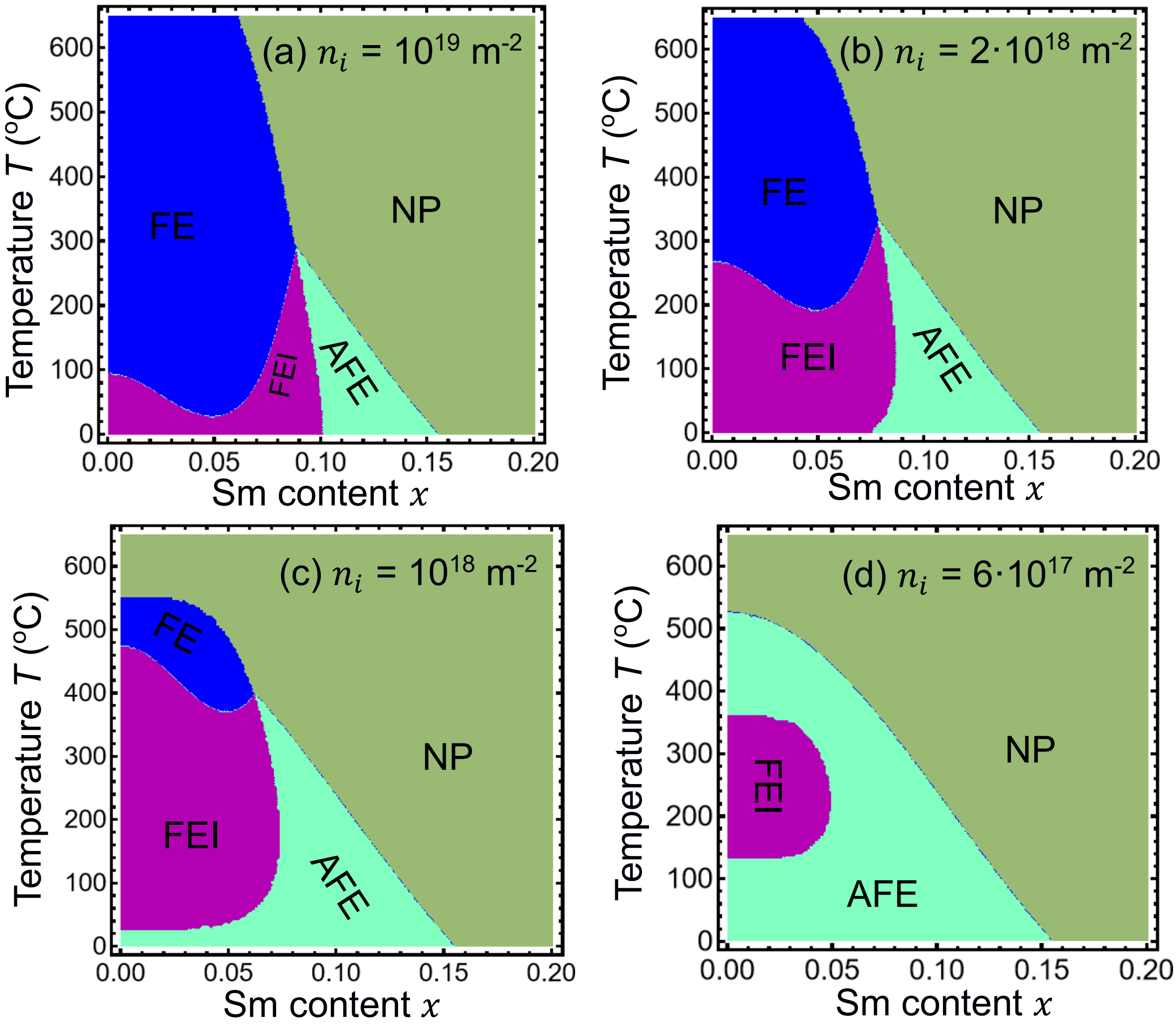

**FIGURE 7.** Phase diagrams of $Bi_{1-x}Sm_xFeO_3$ nanoparticles in dependence on the Sm content $x$ and temperature $T$. The diagrams are calculated using the free energy (3) for the same parameters as in **Fig. 3,** $R = 5$ nm and $\varepsilon_M = 10$ and several values of $n_i = 10^{19}$m$^{-2}$ **(a)**, $2 \cdot 10^{18}$m$^{-2}$ **(b)**, $10^{18}$m$^{-2}$ **(c)** and $6 \cdot 10^{17}$m$^{-2}$ **(d)**.

The influence of the surface charge density on the reentrant transition from the FE phase to the FEI phase is very strong entire the actual range of $n_i$ variation from $10^{17}$m$^{-2}$ to $10^{20}$m$^{-2}$ (as shown in **Fig. 8(a)** for the room temperature, $R = 5$ nm and $\varepsilon_M = 30$). The influence of $n_i$ on the transition from the FEI phase to the AFE phase is a little weaker than the influence of $n_i$ on the FE-FEI transition. The influence of $n_i$ on the AFE-NP transition is absent, because the depolarization field is absent in the homogeneous AFE and NP phases. The ultra-thin orange stripe near the FEI-AFE boundary corresponds to the poly-domain FEI phase.

Dependencies of the spontaneous polarization $P_S$ and antipolar order parameter $A_S$ on the Sm content $x$ and the ionic-electronic charge density $n_i$ are shown in **Fig. 8(b)** and **Fig. 8(c)**, respectively. Similarly to the situation shown in **Fig. 4(b)** and **6(b)**, $P_S$ does not have any features at the FEI-FE phase boundary; decreases significantly approaching the AFE boundary and disappears in the AFE phase. Inside the regions of FE and FEI phases $P_S$ monotonically increases with increase in $n_i$ and saturates for high $n_i$ values. At $n_i < n_{cr}$ the transition from the FEI phase to the AFE phase occurs due to the increase of depolarization field under the weak screening by the surface charge. Note that $n_{cr} \approx 3.14 \cdot 10^{17}$ m$^{-2}$ at $R = 5$ nm and $\varepsilon_M = 30$. $P_S$ is maximal for $x = 0$, decreases monotonically with $x$ increase and eventually disappears for $x \geq x_{cr}(n_i)$, where $x_{cr}$ increases with increase in $n_i$ and then saturates to the "bulk" value $x_C = 0.1$.

Similarly to the situation shown in **Fig. 4(c)** and **6(c)**, the antipolar order parameter $A_S$ exists in the FEI and AFE phases. It gradually disappears approaching the FEI-FE boundary (which shape is close to the tilted parabola), as well as approaching the AFE-NP boundary (which is close to the vertical line $x \approx x_A = 0.15$ independently on $R$ and $\varepsilon_M$ values). $A_S$ has a diffuse maximum at the FEI-AFE boundary and reaches absolute maximum in a relatively big region of the AFE phase with $n_i < n_{cr}$ and $x < x_{cr}$. $A_S$ is maximal in this region, because the weak screening suppresses the spontaneous polarization.

The dependences of the nanoparticle dielectric susceptibility on the Sm content $x$ calculated at $E = 0$ for several values of the ionic-electronic charge density $n_i$ are shown in **Fig. 8(d)**. For $n_i < n_{cr}$ the susceptibility monotonically decreases with increase in $x$ (see the black curve in **Fig. 8(d)**), because the FEI-AFE transition is absent for the range of charge densities in **Fig. 8(a)**. For $n_i \gg n_{cr}$ the susceptibility diverges at the critical value of Sm content, $x = x_{cr}$, that increases with increase in $n_i$ (compare red, green and blue curves in **Fig. 8(d)**). The divergency corresponds to the second order transition between the FEI and AFE phases. Blue arrows in **Fig. 8(d)** point on the hardly visible tiny fractures, which correspond to the reentrance of the FEI phase for the highest value of $n_i = 10^{19}$ m$^{-2}$. For $n_i < n_R^{\min}$ any features of the reentrant FE-FEI transition are not visible on the dielectric susceptibility curve, since the transition is absent for the charge density range (see **Fig. 8(a)**). Note that $n_R^{\min} \approx 3 \cdot 10^{18}$ m$^{-2}$ for $R$ =5 nm and $\varepsilon_M = 30$. The orange curve in **Fig. 8(d)** corresponds to $n_i = 3.3 \cdot 10^{17}$m$^{-2}$ that is slightly higher $n_{cr}$. The susceptibility curve contains a vertical jump at the boundary between the single-domain and poly-domain FEI states, and the sharp maximum at the poly-domain FEI – AFE boundary.

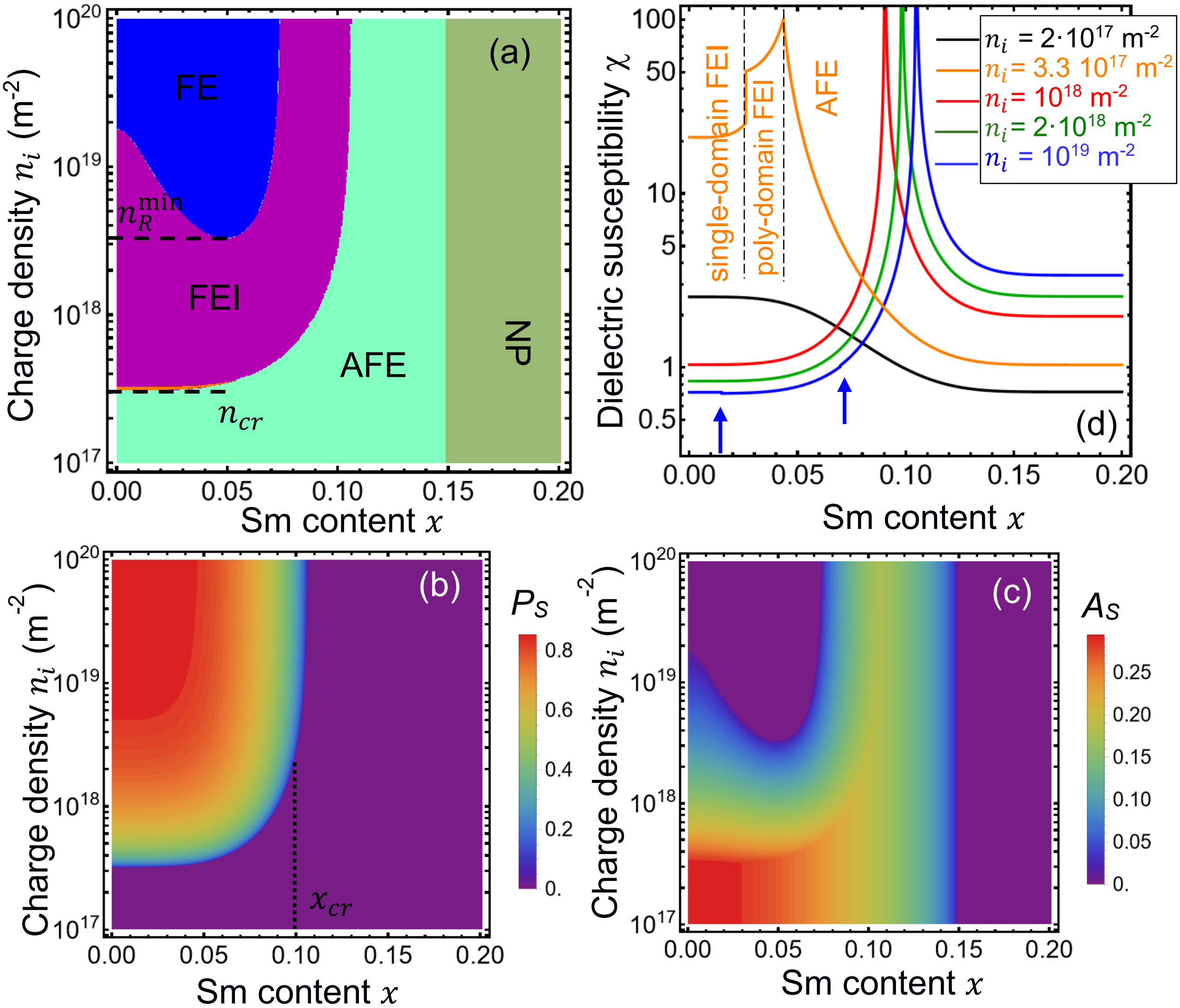

**FIGURE 8.** **(a)** The phase diagram of $Bi_{1-x}Sm_xFeO_3$ nanoparticles in dependence on the Sm content $x$ and ionic-electronic charge density $n_i$. Dependencies of the spontaneous polarization $P_S$ **(b)** and antipolar order $A_S$ **(c)** on x and $n_i$ **(d)** The dependence of the nanoparticle dielectric susceptibility on the Sm content $x$

calculated at $E = 0$ for $n_i = 2 \cdot 10^{17}\text{m}^{-2}$ (the black curve), $3.3 \cdot 10^{17}\text{m}^{-2}$ (the orange curve), $10^{18}\text{m}^{-2}$ (the red curve), $2 \cdot 10^{18}\text{m}^{-2}$ (the green curve) and $10^{19}\text{m}^{-2}$ (the blue curve). Blue arrows point on the tiny fractures, which correspond to the reentrance of the FEI phase. The temperature $T$ =25ºC; $R = 5$ nm and $\varepsilon_M$ =30, other parameters are the same as in **Fig. 3.**

Hence, the influence of the ionic-electronic charge density $n_i$ on the spontaneous polarization is the most significant near the FEI-AFE boundary. The influence of $n_i$ on the antipolar order is the most significant below the "bottom" of the reentrant FEI-FE boundary, as well as below the bottom of the FEI-AFE boundary (i.e., for $n_i \leq 3 \cdot 10^{17}\text{m}^{-2}$ and $x < 0.05$ at $R$ =5 nm and $\varepsilon_M$ =30).

As it follows from **Figs. 3 – 8**, it is possible to control the transition from the FE phase to the FEI phase and then to the AFE by decreasing the size of the $Bi_{1-x}Sm_xFeO_3$ nanoparticle, or by decreasing the dielectric permittivity of the environment, or by decreasing the density of the surface ionic-electronic charge. From comparison of **Figs. 4**, **6** and **8** we conclude that the influence of the surface ionic-electronic charge on the phase state of the $Bi_{1-x}Sm_xFeO_3$ nanoparticles is the strongest. The possibility to change the surface charge density is the most interesting and promising for applications, because the change of $n_i$ allows to control the phase diagrams of the nanoparticles by ions adsorption from the environment. At the same time, the surface charge injection should change of the field-induced polar properties of the $Bi_{1-x}Sm_xFeO_3$ nanoparticles.

The quasi-static dependences, $P_3(E_3)$, calculated for different Sm content $x$ in the $Bi_{1-x}Sm_xFeO_3$ nanoparticles, are shown in **Fig. A1** in **Appendix A** [32] for the case of the relatively small ($\xi \ll 1$) and big ($\xi \gg 1$) coupling constants $\xi$. For $\xi \ll 1$ the decrease in $n_i$ leads to the degradation of ferroelectric loop, its eventual disappearance and appearance of the paraelectric-like response of polarization to the external field. For $\xi \gg 1$ the quasi-rectangular ferroelectric single loops at first transforms to the antiferroelectric-type double-loops and then to the linear hysteresis-less curves with $n_i$. The further decrease of $n_i$ leads to the gradual shrinking of the double loops, which eventually transform in the linear dielectric-like polarization response to the external field.

For $\xi \ll 1$ and relatively high density of surface ions ($n_i{\sim}10^{18}\text{m}^{-2}$) the increase of Sm content $x$ from 0 to 0.10 leads to the gradual transition of the quasi-rectangular ferroelectric-type single loops of the polarization to the hysteresis-less paraelectric curves (see **Figs. A2** in **Appendix A** [32]).

### B.4. The influence of the ferro-ionic coupling on the domain structure of $Bi_{1-x}Sm_xFeO_3$ nanoparticles

As it was mentioned earlier, the relatively small densities $n_i \leq 10^{17}$m$^{-2}$ cannot provide a sufficient screening of the nanoparticle spontaneous polarization (see e.g., **Fig. 2(c)**). In result the single-domain ferroelectric state has higher free energy in comparison with the poly-domain state. When the energy of the polydomain state is negative (and thus lower than the zero energy of the NP state) the appearance of the polydomain structures becomes energetically preferable.

Hypothetically, the diagrams shown in **Figs. 3-6** and **7(a)** - **7(c)**) may contain the region of the poly-domain FE and FEI states. To verify the possibility, we used the direct variational method [45], which allows us to consider possible domain formation in the harmonic approximation and checked obtained results by FEM. Appeared that the possibility is not realized because the concentrations $n_i \geq 10^{18}$m$^{-2}$ are high enough to prevent the domain formation for other parameters fixed in the figures. The single-domain and poly-domain states may coexist inside the FEI phase region in **Fig.7(d)** calculated for $n_i < 10^{18}$m$^{-2}$, however the effect of the "geometric catastrophe" suppresses the domain formation in small nanoparticles (e.g., for 10 nm size or smaller). A very thin region of the poly-domain FEI state (corresponding to $n_i$ slightly higher $n_{cr}$) exists in **Fig. 8(a)**.

Considering that a given domain structure is characterized by polarization contrast and amplitude, period and/or other features of domain morphology, the changes of the domain structure with Sm content $x$ and $n_i$ are shown **Figs. A3** in **Appendix A** [32]. The patterns are calculated by the FEM for the $T$ =25°C, $R$ = 20 nm and $\varepsilon_M$ =30.

Typical scenario of the emergence and changes of the domain structure morphology, which appear with increase in $n_i$, are shown in **Fig. 9** for $R$ =20 nm. For $n_i \leq 5 \cdot 10^{16}$ m$^{-2}$ the fine and faint ferroelectric domain stripes appear inside the nanoparticle (see the top image in **Fig. 9(a)**). It is seen from the figure that the ferroelectric domain walls induce the modulation of the antipolar order (see the bottom image in **Fig. 9(a)**). The stripes period and contrast increase with increase in $n_i$ (see the top image in **Fig. 9(b)**), at that the modulation of the antipolar order becomes much more pronounced (see the bottom image in **Fig. 9(b)**). For $n_i \geq 5 \cdot 10^{17}$ m$^{-2}$ the screening-assisted transition to the single-domain state occurs (see **Fig. 9(c)**).

The phase diagram of the 40-nm $Bi_{1-x}Sm_xFeO_3$ nanoparticles in dependence on the Sm content $x$ and ionic-electronic charge density $n_i$ is shown **Fig. 9(d)**. In contrast to the case of the 10-nm nanoparticle diagram (shown in **Fig. 8(a)**), the large region of the poly-domain FEI phase exists for $n_i < (1.5-3) \cdot 10^{17}$m$^{-2}$ and $x < 0.09$. The susceptibility curves calculated at $E = 0$ for $10^{16}\mathrm{m}^{-2} < n_i < 1.5 \cdot 10^{17}\mathrm{m}^{-2}$ contain the sharp maximum at the boundary between the poly-domain FEI states and the AFE phase (see the black curve in **Fig. 8(e)**). The susceptibility curves calculated for $1.5 \cdot 10^{17}\mathrm{m}^{-2} < n_i < 3 \cdot 10^{17}\mathrm{m}^{-2}$ contain a pronounced vertical jump at the boundary between the single-domain and poly-domain FEI states, and the sharp maximum at the poly-domain FEI – AFE boundary (see red and green curves in **Fig. 8(e)**). The susceptibility curves calculated for $n_i > 1.5 \cdot 10^{17}\mathrm{m}^{-2}$ contain the divergency at the boundary between the single-domain FEI states and the AFE phase (see the blue curve in **Fig. 8(e)**).

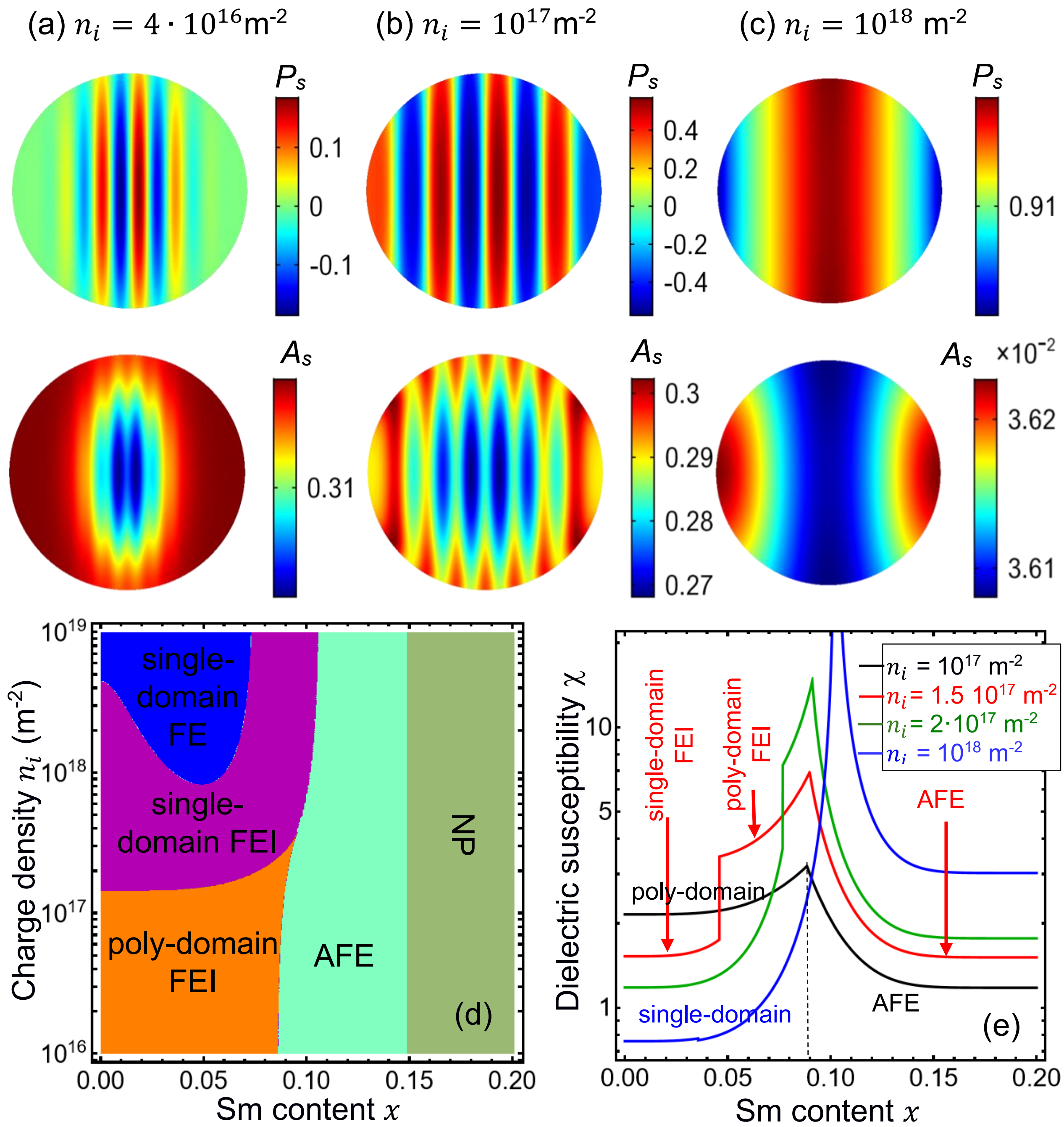

**FIGURE 9.** Typical view of the equilibrium ferroelectric polar (the top row) and antipolar (the middle row) domain structures calculated by the FEM in the $Bi_{0.95}Sm_{0.05}FeO_3$ nanoparticle covered with the ionic-electronic screening charges with the surface density $n_i = 4 \cdot 10^{16}$m$^{-2}$ **(a)**, $10^{17}$m$^{-2}$ **(b)**, and $10^{18}$m$^{-2}$ **(c)**. **(d)** The phase diagram of $Bi_{1-x}Sm_xFeO_3$ nanoparticles in dependence on the Sm content $x$ and ionic-electronic charge density $n_i$. **(e)** The dependence of the $Bi_{1-x}Sm_xFeO_3$ nanoparticle dielectric susceptibility on the Sm content $x$ calculated at $E = 0$ for $n_i = 10^{17}$m$^{-2}$ (the black curve), $1.5 \cdot 10^{17}$m$^{-2}$ (the red curve), $2 \cdot 10^{17}$m$^{-2}$ (the green curve) and $10^{18}$m$^{-2}$ (the blue curve). Parameters $T$ =25ºC, $R$ = 20 nm and $\varepsilon_M$ =30; other parameters are the same as in **Fig. 3.**

Note that the appearance, re-building or disappearance of the domain structure, which takes place in the definite range of $x$ and/or $n_i$, should lead to the changes in the nanoparticle dielectric susceptibility (compare the poly-domain black curve, "mixed" red and green curves, and the single-domain blue curve in **Fig. 9(e)**). The features related with the domain structure remain visible in the presence of small external field and smear with the field increase (compare different curves in **Fig. A4** in **Appendix A** [32]).

## III. EXPERIMENTAL RESULTS

It was shown a while ago that the electronic properties (e.g., the band gap) of Sm-doped $BiFeO_3$ nanoparticles can be regulated by the Sm content [46]. It has been shown recently that the ferroelectric and/or multiferroic properties of $BiFeO_3$ nanoparticles can be controlled by co-doping with $Sm^{3+}$ and $Co^{2+}$ ions [47] and/or by the facile synthetic parameters [48].

The substitution of $Bi^{3+}$ ions by $Cd^{3+}$ or $Ho^{3+}$ ions up to 15% in $BiFeO_3$ nanoparticles synthesized by sol-gel method are studied in detail in Refs.[49] and [50], respectively. In particular, the substitution-driven structural phase transition from the rhombohedral (*R3c*) to the orthorhombic (*Pnma*) phase has been observed in the $Bi_{1-x}Gd_xFeO_3$ nanoparticles with increase in $x$ up to 15 % [49]. Also, a weak ferromagnetic state has been observed and attributed to the partial suppression of spiral spin structure, surface effect and Dzyaloshinskii-Moriya spin-orbit interactions. The Rietveld refinement analysis of X-ray diffraction profiles reveals the formation of a pure rhombohedral phase with nanocrystalline nature in $Bi_{1-x}Ho_xFeO_3$ nanoparticles with $x = 0.0 - 0.05$, whereas the mixed rhombohedral + orthorhombic state is obtained for $x = 0.10$ [50]. The $Ho^{3+}$ substitution can induce the transformation to the orthorhombic phase for $x = 0.15$. The dielectric properties, such as dielectric constant and loss factor, decrease with increasing $Ho^{3+}$ content.

A series of nanopowder samples of $Bi_{1-x}Sm_xFeO_3$ ($0 \leq x \leq 0.2$) were synthesized using the solution combustion method (see **Fig. 10(a)**). Preparation details are given in **Appendix B** [32]. The synthesized samples were calcined for 5 hours at 750 °C to minimize the presence of residual water and hydroxyl groups. The obtained six samples were named as the BFO, SFO, BSFO-005, BSFO-010, BSFO-015 and BSFO-020, respectively. According to the X-ray diffraction (XRD) data, summarized in Table **B1** in **Appendix B** [32], the BFO sample contains 73 % of the long-range ordered rhombohedral (*R*3c) $BiFeO_3$ phase, 16 % of the orthorhombic (*Pbam*) $Bi_2Fe_4O_9$ phase and 11% of the cubic (*I23*) $Bi_{25}FeO_{40}$ phase. The Sm-doping very strongly increases the phase purity of the nanopowders. The BSFO-005 contains 97% of the $Bi_{0.95}Sm_{0.05}FeO_3$ in the *R*3c phase, 2 % of the $Bi_2Fe_4O_9$ and 1% of the $Bi_{25}FeO_{40}$. The BSFO-010 contains 99% of the $Bi_{0.9}Sm_{0.1}FeO_3$ in the *R*3c phase and 1% of the $Bi_{25}FeO_{40}$. The BSFO-015 contains 7 % and 92 % of the $Bi_{0.85}Sm_{0.15}FeO_3$ in the polar *R*3c and orthorhombic *Pbnm* phases, respectively, and 1% of the $Bi_{25}FeO_{40}$. The BSFO-020 sample contains 99 % of the $Bi_{0.8}Sm_{0.2}FeO_3$ in the *Pbnm* phase and 1% of the $Bi_{25}FeO_{40}$. The SFO sample contains 100 % of the $SmFeO_3$ in the *Pbnm* phase. Since the polar or/and antipolar long-range order can exist (or coexist) in the *R*3c phase only, the results of the XRD phase analysis are in quantitative agreement with the phase diagrams of $Bi_{1-x}Sm_xFeO_3$ nanoparticles calculated in the previous section. Indeed, the *R*3c phase completely disappears in the $Bi_{1-x}Sm_xFeO_3$ nanoparticles for $x \geq 0.15$ in accordance with XRD analysis and analytical calculations, which predict the transition to the NP phase at $x \approx 0.15$. The phase transformations, observed in $Bi_{1-x}Sm_xFeO_3$ nanoparticles with increase in $x$ from 0 to 0.15, correlate with the phase transformations observed in $Bi_{1-x}Gd_xFeO_3$ nanoparticles [49] and $Bi_{1-x}Ho_xFeO_3$ nanoparticles [50] with increase in Gd (or Ho) content $x$ from 0 to 0.15.

Since the size and surface effects can play a key role in identifying the physical properties of the $Bi_{1-x}Sm_xFeO_3$ nanoparticles, it is important to determine the actual sizes and surface morphology, which can be responsible for the properties' manifestation. Transmission electron microscopy (TEM) shows the presence of very small (size less than 10 nm) and very big (size more than 200 – 300 nm) $Bi_{1-x}Sm_xFeO_3$ nanoparticles (see typical TEM images in **Fig. 10(b)-9(f)**). The Rietveld analysis of the XRD data shows that the substitution of Bi ions by Sm ions is accompanied by a decrease in the size of the coherent scattering regions (CSR). The CSR size was determined from the broadening of the XRD peaks (shown in **Fig. B1** in **Appendix B** [32]). The size of the CSR is approximately 103 nm for $x = 0$, decreases to 55 nm for $x = 0.10$ and decreases further up to 46 nm for $x = 1$. Since the sizes of the CSR are the same order as the period of the magnetic cycloid in a pure $BiFeO_3$ (about 60 nm), it is reasonable to conclude that the CSR size effect plays the key role in the formation of the magnetoelectric response of the $Bi_{1-x}Sm_xFeO_3$ nanoparticles.

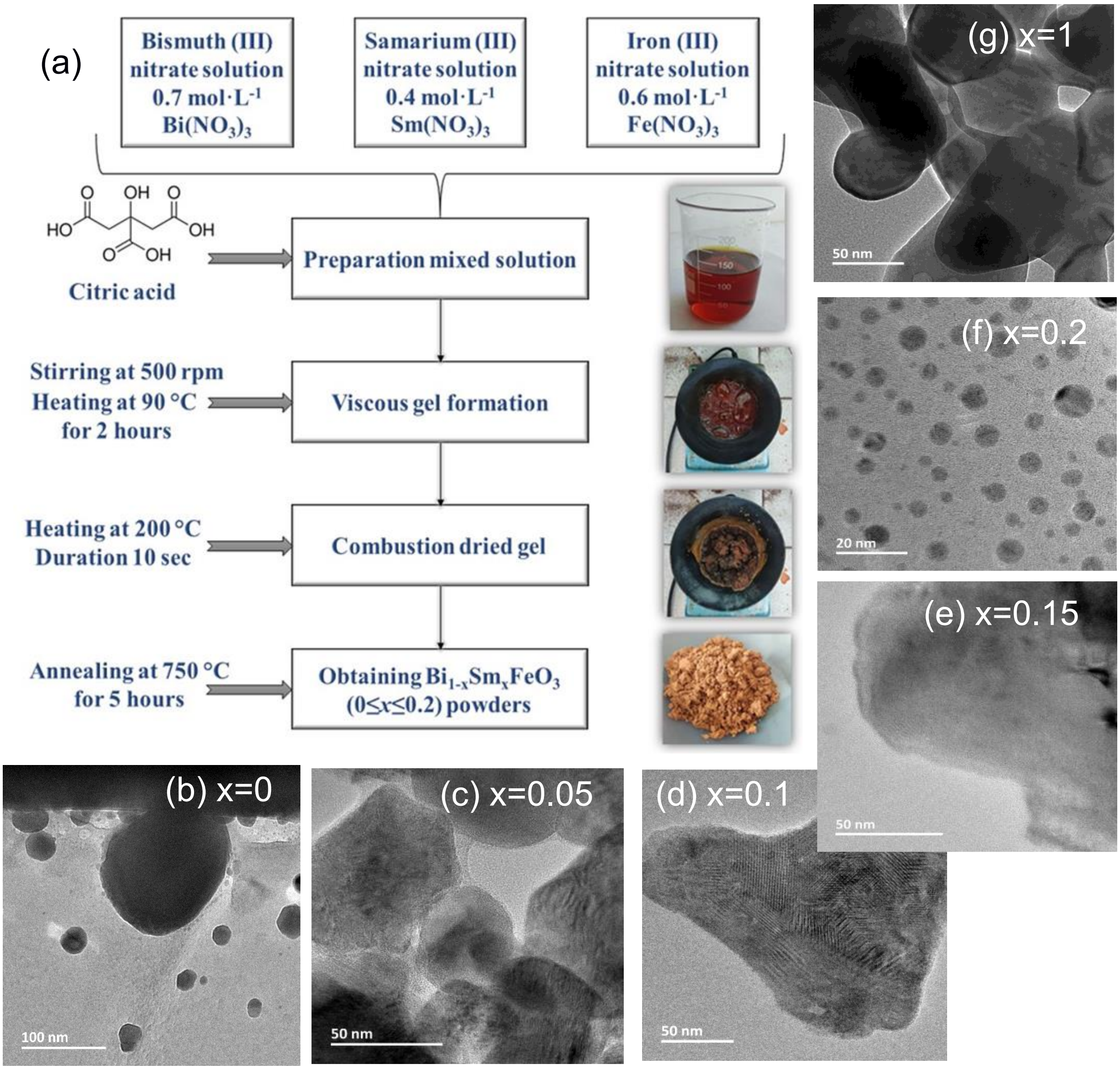

**FIGURE 10. (a)** Schematic illustration of the solution combustion process to prepare the $Bi_{1-x}Sm_xFeO_3$ ($0 \le x \le 0.2$) nanopowders**. (b)-(g)** Typical TEM images of $Bi_{1-x}Sm_xFeO_3$ nanoparticles.

Next, the $Bi_{1-x}Sm_xFeO_3$ nanopowders were characterized by the electron paramagnetic resonance (EPR) and Fourier-transform infra-red (FTIR) spectroscopies, and electrophysical measurements. Obtained results are discussed below.

### A. Electron paramagnetic resonance spectroscopy

EPR measurements were carried out on X-band Bruker Elexsys E580 spectrometer operating at 9.8 GHz frequency at room temperature. The EPR spectra of the as-prepared samples of $Bi_{1-x}Sm_xFeO_3$ nanoparticles with increasing Sm content ($x$ =0, 0.1, 0.15, 0.2 and 1) are presented in **Fig. 11.** Two EPR lines with the effective values of $g$-factor near 2 and 6 can be observed. The value of $g_{eff}$ for these two lines and their linewidth, calculated from the peak-to-peak distance, are summarized in **Table I.** The value of $g_{eff}$ was calculated using the equation, $g = \hbar\omega/(\beta H_r)$, where $\hbar$ is Planck's constant, $\omega$ is the operating circular frequency, $\beta$ is the Bohr magneton, and $H_r$ is the resonant magnetic field.

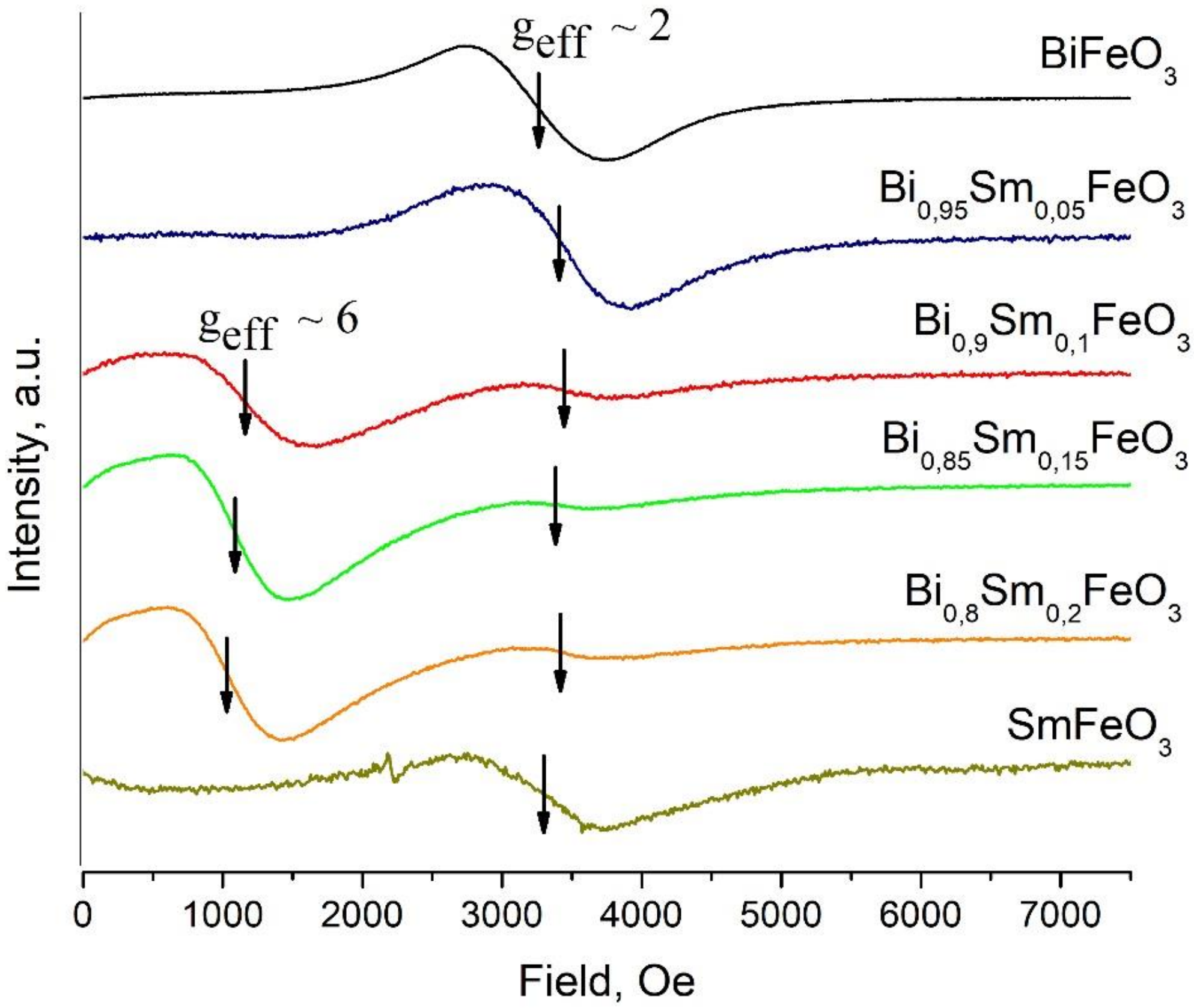

**FIGURE 11.** EPR spectra of $Bi_{1-x}Sm_xFeO_3$ nanopowders measured at 300 K.

**Table I**. EPR parameters calculated for $Bi_{1-x}Sm_xFeO_3$ nanoparticles

| Sample | $g_{eff}$ | Linewidth (ΔH, Oe) | $\Delta g/g$ | $g_{eff}$ | Linewidth (ΔH, Oe) |
|---|---|---|---|---|---|
| $BiFeO_3$ | 2.190 | 860 | 0.094 | - | - |
| $Bi_{0.95}Sm_{0.05}FeO_3$ | 2.060 | 1020 | 0.029 | - | - |
| $Bi_{0.9}Sm_{0.1}FeO_3$ | 2.02 | 562 | 0.009 | 6.2 | 1100 |
| $Bi_{0.85}Sm_{0.15}FeO_3$ | 2.05 | 460 | 0.024 | 6.5 | 850 |
| $Bi_{0.8}Sm_{0.2}FeO_3$ | 2.025 | 520 | 0.011 | 6.8 | 870 |
| $SmFeO_3$ | 2.17 | | 0.09 | - | - |

The EPR spectrum of pristine $BiFeO_3$ consists of one wide line at $g_{eff} \approx 2.19$ being in good agreement with the literature data [51, 52, 53] for this compound. The spectra of all samples are broad and rather asymmetric, which indicate the presence of strong magnetic interaction in the samples.

The magnetic properties of $BiFeO_3$ are determined by the $Fe^{3+}$ ions, which experience the super-exchange interaction via intermediate oxygen atoms, resulting in the G-type antiferromagnetic ordering that persists up to 643 K [54]. Due to the antisymmetric Dzyaloshinskii-Moriya spin-orbit interaction, the magnetic moments on the Fe atoms are canted from their antiparallel directions and form the incommensurate spin cycloid structure superimposed on the AFM order. The effect results in the contribution of the Fe sublattice in the electric polarization due to the tilt of the oxygen octahedra responsible for both polar and antipolar order parameters and links the magnetic and ferroelectric orders in $BiFeO_3$ [55, 56, 57].

EPR experiments performed at different temperatures and magnetic fields have attributed the wide line with $g_{eff} \approx 2$ to resonant absorption in the cycloidal spin structure (see e.g. Refs.[53, 58]). Partial substitution of bismuth by samarium shifts the $g_{eff}$ magnitude of this line toward smaller values and leads to the formation of an additional broad line with $g_{eff} \approx 6$ for $x \geq 0.1$.

The magnitude of $g_{eff}$ can be used to estimate the spin canting of $Fe^{3+}$ ions and the tilt of the oxygen octahedra using the vector parameter $D \approx \left(\frac{\Delta g}{g}\right) J_{super}$, where $\Delta g$ is deviation of $g$ from 2.0023, and $J_{super}$ corresponds to the super-exchange interaction coefficient [51, 52]. Assuming that $J_{super}$ does not change for the $Fe^{3+}$ ions belonging to the spin cycloid, the monotonic decrease in the $D$ value for $x$ from 0 to 0.1 (see **Table I**) indicates a decrease in the tilt of the octahedra, and, according to the calculated phase diagrams (see e.g., **Fig. 3(a)**), a decrease in the polarization of nanopowders during the FEI-AFE transition, which occurs near $x = 0.1$. The change in the tilt of the oxygen octahedra at $x = 0.15$ reflects the AFE-NP transition (see e.g., **Fig. 3(a)** and **4(c)**).

It should be noted that such behavior of the g-factor contradicts the previously obtained data for $Bi_{1-x}Sm_xFeO_3$ nanoparticles [52]. Moreover, to the best of our knowledge, the previous EPR spectra recorded for the $Bi_{1-x}Sm_xFeO_3$ compounds show only one line with the $g$-factor near 2. This may indicate the formation of parasitic phases, but the addition of samarium suppress the formation of the secondary phases in $BiFeO_3$ (see XRD data and e.g., Ref.[31]). The difference can be the sequence of the size effect, when the role of the nanoparticle surface begins to dominate over the role of its volume in the EPR spectrum formed by the local EPR-active centers.

The line with the value $g_{eff} \approx 6$ cannot be assigned to the absorption in the spin cycloid, but is typical for an EPR signal originating from isolated $Fe^{3+}$ ions located in crystal field with distorted oxygen-coordinated environments (see Ref.[54] and **Appendix B** in Ref.[32]).

Considering the presence of the G-type antiferromagnetic ordering in $BiFeO_3$, the origin of the EPR signal at $g_{eff} \approx 6$ can be explained by the violation of the antiferromagnetic ordering by oxygen vacancies or other defects presented in the studied samples. This results in formation of isolated $Fe^{3+}$ ions in distorted octahedral environment, which experience the dipole-dipole and super-exchange interaction leading to a broad line in the EPR spectrum. The intensity of this line non-monotonously increases with increase of Sm content. The accompanying increase in $g_{eff}$ from 6.2 for $x = 0.1$ to 6.8 for $x = 0.2$ indicates an increase in the local magnetic field at the $Fe^{3+}$ ions cites and is associated with the formation of superparamagnetic clusters due to the break of the antiferromagnetic ordering by oxygen vacancies. According to the X-ray diffraction (XRD) data, shown in **Fig. B1**, substitution of bismuth by samarium atoms is accompanied by a decrease in particle size, which leads to an increase in the number of defects in samples with a large amount of samarium and therefore to a higher intensity of the EPR signal at $g_{eff} \approx 6$. In this case, the presence of the large number of oxygen vacancies affects the value of $J_{super}$, which does not allow the use of the $D$ value to estimate the tilt of the oxygen octahedra.

To summarize, the shape of the EPR spectra of the studied $Bi_{1-x}Sm_xFeO_3$ nanoparticles undergoes three distinctive changes at $x = 0.10$ ($\Delta g/g$ for the line with $g_{eff} \approx 2$ reaches its minimum and the line with $g_{eff} \approx 6$ appears), at $x = 0.15$ (according to $\Delta g/g$ behavior), and at $x > 0.2$ (according to $\Delta g/g$ behavior the line with $g_{eff} \approx 6$ disappears). These results are in a qualitative agreement with the phase diagram shown in **Fig. 3(a)**, which contains the FEI-AFE transition near $x = 0.1$ and AFE-NP transition at $x = 0.15$. In particular, the dashed horizontal line corresponding to the temperature 27°C = 300 K in **Fig. 3(a)** crosses the FEI-AFE boundary at $x \approx 0.105$ and the AFE-NP boundary at $x \approx 0.150$.

### B. Fourier-transform infra-red spectroscopy of the $Bi_{1-x}Sm_xFeO_3$ nanoparticles

Typical FTIR spectra in the region (1100 – 400) $cm^{-1}$ for $Bi_{1-x}Sm_xFeO_3$ ($x = 0.05 - 0.2$) and $SmFeO_3$ nanopowder samples calcined at 750°C are shown in **Fig. 12.** The FTIR spectrum of the $BiFeO_3$ nanopowder is characterized by the bands at 445 and 540 $cm^{-1}$, which correspond to the stretching and bending vibrations of the Fe–O bonds, respectively, and are typical for $FeO_6$ octahedra in perovskites [59]. The bands near 810 and 1070 $cm^{-1}$ are attributed to the Bi–O bond vibrations [60, 61].

When $Bi^{3+}$ ions (with ionic radius 1.365 Å) are partially substituted with smaller $Sm^{3+}$ ions (ionic radius 1.132 Å), the changes occur in the unit cell parameters and the symmetry can also change. In particular, the decrease in the Fe–O–Fe bond angle should be reflected in the FTIR spectra and allows tracking the modification of the vibration modes belonging to different structural phases. Strong peaks at ~440 and ~550 $cm^{-1}$ correspond to the stretching and bending vibrations, respectively, occurring in the $FeO_6$ octahedra in the rhombohedral structure [59, 62]. With increasing the Sm content, the position of the Fe-O bond shifts towards higher wavenumbers, which may be due to the distortion of the $FeO_6$ octahedra. In the nanopowders with 15 % of Sm content, a band at 680 $cm^{-1}$ is observed, which may be related to the modification of the dipole moment typical for the anti-polar orthorhombic structure. Such spectral changes may be associated with changes in the phase state of the samples, indicating on the possible phase transition from the rhombohedral (space group $R3c$) structure to the orthorhombic (*Pbnm* space group) structure [63]. From the theoretical standpoint, the band at 680 $cm^{-1}$ may be related to the AFE-NP transition, when the antiparallel orientation of the dipole moments disappears in zero or very small electric field (see the behavior of the spontaneous antipolar order $A_S$ in **Fig. 4(c)**).

Taking into account the phase diagrams calculated for 100 - 20 nm nanoparticles, where the orthorhombic AFE phase can be stable at room temperature for Sm content $0.09 < x < 0.16$ and the NP phase becomes stable at higher Sm content $x \geq 0.16$ (see e.g., **Figs. 3(a)**, **3(b)**, **5(a)**, **5(b)**, **7(a)** and **9(d)**), and XRD results show the dominant (more than 90 %) orthorhombic *Pbnm* phase for $x \geq 0.15$ (see **Appendix B** in Ref.[32]), we can assume that the band at 680 $cm^{-1}$ is possible for the nanopowders with Sm content $0.1 < x < 0.2$. The band at 680 $cm^{-1}$ disappears with further increase of Sm content up to 20 %, and the position of the Fe-O bond band shifts again to the region of lower wavenumbers, which indicates the completed phase transition with a corresponding change in the unit cell parameters, stabilization of the nonpolar orthorhombic phase and characterizes deformation vibrations of the Fe-O bond in the nonpolar orthorhombic structure. Since the theory gives a narrower range of the AFE phase stability, mainly $0.1 < x \leq 0.16$, the XRD experiment gives that the domination of the *Pbnm* phase starts $x \geq 0.15$, and FTIR gives $0.1 < x < 0.2$, we can assume the AFE-NP phase transition in the studied nanopowders can occur in the range of Sm content $0.15 < x < 0.2$.

The FTIR spectrum of the $SmFeO_3$ nanopowder is characterized by two asymmetric broad bands around 414 and 555 $cm^{-1}$, indicating overlapping bands that belong to the stretching and bending vibrations of Fe–O and Sm–O, which is consistent with the literature [64].

Thus, based on the analysis of the FTIR spectra, we can conclude that the complete phase transition to the NP orthorhombic phase occurs when (16 – 20) % of $Bi^{3+}$ ions are substituted with $Sm^{3+}$ ions, which agrees with the AFE-NP phase transition calculated for the $Bi_{1-x}Sm_xFeO_3$ nanoparticles (see e.g., **Figs. 3(a)**, **3(b), 4(c), 5(a)**, **5(b), 7(a)** and **9(d)**).

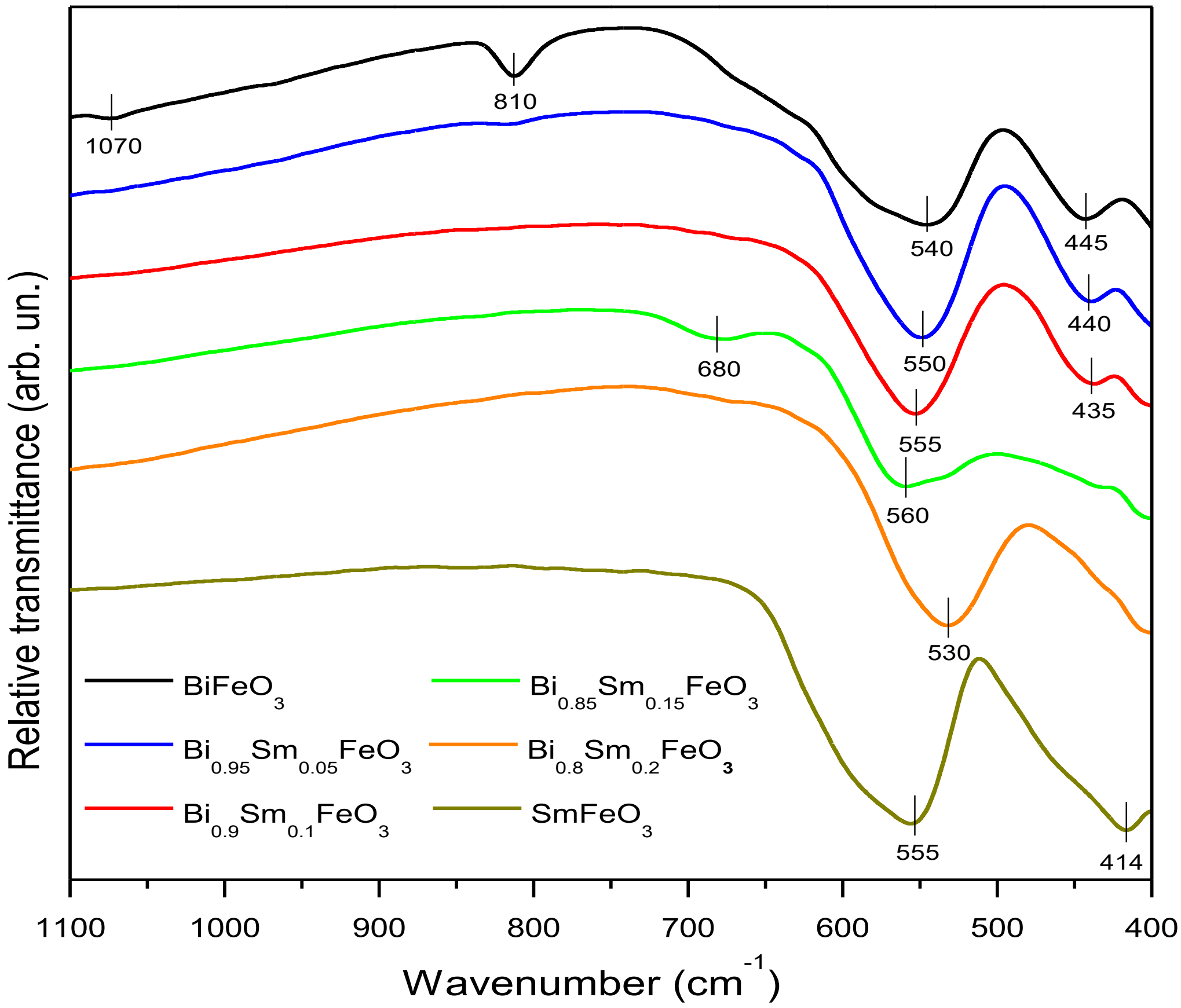

**FIGURE 12.** FTIR spectra of $Bi_{1-x}Sm_xFeO_3$ nanopowders measured at room temperature.

### C. Results of electrophysical measurements

To perform electrophysical and dielectric measurements, the $Bi_{1-x}Sm_xFeO_3$ nanopowders were pressed in Teflon cells between two metallic plungers, which serve as electric contacts (see insets in **Fig. 13**). The samples in the cell have a disk shape with the 6 mm diameter and (0.01 – 0.05) mm thickness. To minimize the influence of the spatial inhomogeneity of the pressed powders on the electrophysical and dielectric measurements, the area of the electrodes was maximized. To avoid the influence of geometry effect on the measured capacitance and losses, all samples in the series have the same thickness, much smaller than their 6 mm diameter. We carefully checked that the application of bigger pressure applied to thinner samples does not lead to any noticeable changes in their electrophysical properties, as well as the measured values do not change after the pressure removal. Note that the compression was used to ensure the stability of the powder sample during the I-V curves recording.

The RLC meters UNI-T UT612 and E7-12 were used to measure the capacitance of the cells in the range (100 – $10^6$) Hz. The electric voltage was applied to the sample connected in series with the load resistor from the software-controlled power supply GW Instek PSP-603. The current-voltage (I-V) characteristics were measured by Keithley-2000 digital multimeters, after 100 hours in the high vacuum, and after the for-vacuum pumping and 30 minutes in the Ar atmosphere.

Frequency dependence of the effective dielectric capacitance and losses of the pressed $Bi_{1-x}Sm_xFeO_3$ nanopowders are shown **Fig. 13(a)** and **Fig. 13(b)**, respectively. The dependences of the effective capacitance on the Sm content $x$ measured at different frequencies of applied bias is shown in **Fig. 13(c)**. Error bars in **Fig. 13(a)** are almost the same as the height of corresponding symbols. Error bars in **Fig. 13(b)** and **Fig. 13(c)** are shown only for $SmFeO_3$ nanopowder, for other $Bi_{1-x}Sm_xFeO_3$ nanopowders the measurable error is smaller than the height of corresponding symbols. Taking into account that the capacitance value is shown in logarithmic scale, the error does not exceed (1 – 10) % of the measured values, being significantly smaller that the height of unexpected "dip" in **Fig. 13(c)**.

From **Fig. 13(a),** the capacitance of all samples monotonically decreases with the frequency increase from 100 Hz to 1 MHz. The strongest decrease (in more than 10 times, from 200 pF to 7 pF) is observed for the pressed $SmFeO_3$ nanopowder, the weaker decrease (in about 7 - 4 times, from 50 - 30 pF to 6 - 7 pF) is observed for the Sm content x=0.05 and 0.1. Even a weaker decrease of the capacitance (in about 3 – 2 times) corresponds to the $x = 0$ and 0.15. The weakest decrease of the capacitance (in about 1.7 times) corresponds to the pressed $Bi_{0.8}Sm_{0.2}FeO_3$ nanopowders. The (10 – 1.5) times decrease in the capacitance under the frequency decrease over 4 decades is regarded the slow frequency dispersion characteristic for the wide-gap ferroelectric materials.

From **Fig. 13(b),** the losses of the pressed $Bi_{1-x}Sm_xFeO_3$ nanopowders monotonically decreases with the frequency increase from $10^2$ Hz to 1 MHz. The strongest decrease (in about 100 times) is observed for the pressed $SmFeO_3$ nanopowder, the weaker decrease (in less than 2 times) is observed for the $BiFeO_3$ nanopowder. The losses of the pressed $SmFeO_3$ nanopowder, which tgδ decreases from 2 at 100 Hz to 1 at 1 kHz, are significantly higher than the losses of other samples, which tgδ changes from 1 to 0.1 in the frequency range at 100 Hz – 1 kHz. We regard that the high losses of the pressed $SmFeO_3$ nanopowder can be the reason of its capacitance increase at low frequences due to the Maxwell-Wagner effect [65].

The dependences of the effective capacitance on the Sm content $x$ measured at different frequencies of applied bias is shown in **Fig. 13(c)**. The nonmonotonic dependences, which are measured in ambient conditions, have the pronounced minimum at $x = 0.15$ observed for all frequences from 100 Hz to 1 MHz, and the fracture at $x = 0.1$ observed in the frequency range 100 Hz – 100 kHz, or local maximum observed at $x = 0.1$ for 1 MHz. The observed fracture or maximum at $x = 0.1$ is in a qualitative agreement with the calculated sharp maximum of the dielectric susceptibility near $x = 0.1$, which corresponds to the poly-domain FEI-AFE transition in the $Bi_{1-x}Sm_xFeO_3$ nanoparticles (see **Fig. 9(e)**). The evident reason for the maximum "suppression" (i.e., its height decreases and width increases) in comparison with the calculated sharp maximum is the big scattering of nanoparticles sizes (from several nm to sub-microns) in the studied nanopowder samples (see TEM images in **Fig. 10**), as well as the domain structure appearance inside big and weakly screened nanoparticles. Possible reasons of the nonmonotonic behavior of the effective capacitance can be the dependence of the particle average size $\bar{R}$ on the Sm content, as well as the specific dependence of $n_i$ on $x$. If the dependences $\bar{R}(x)$ and $n_i(x)$ have opposite $x$-trends, the local minima of the effective capacitance can appear. A possible reason for the effective capacitance nonmonotonic behavior and pronounced minimum at $x = 0.15$ (unexplained by the simplified theory presented in this work) may be the morphotropic AFE-NP phase boundary (see e.g., figure 1 in Ref.[2]). The question requires further experimental and theoretical studies.

Typical I-V curves, measured after 100 hours in high vacuum, demonstrate the pronounced loop opening in comparison with the I-V curves measured after the for-vacuum pumping and 30 minutes in Ar atmosphere (compare the black and red curves in **Fig. 13(d)**). This behavior indicates the significant role of the ferro-ionic coupling in the pressed $Bi_{1-x}Sm_xFeO_3$ nanopowders, being in a full agreement with the experimental observations of the strong ferro-ionic coupling in other ferroelectrics in high-vacuum conditions, such as $Hf_xZr_{1-x}O_2$ and $BaTiO_3$ thin films (compare **Fig. 13(d)** with Fig.3 in Ref.[23]).

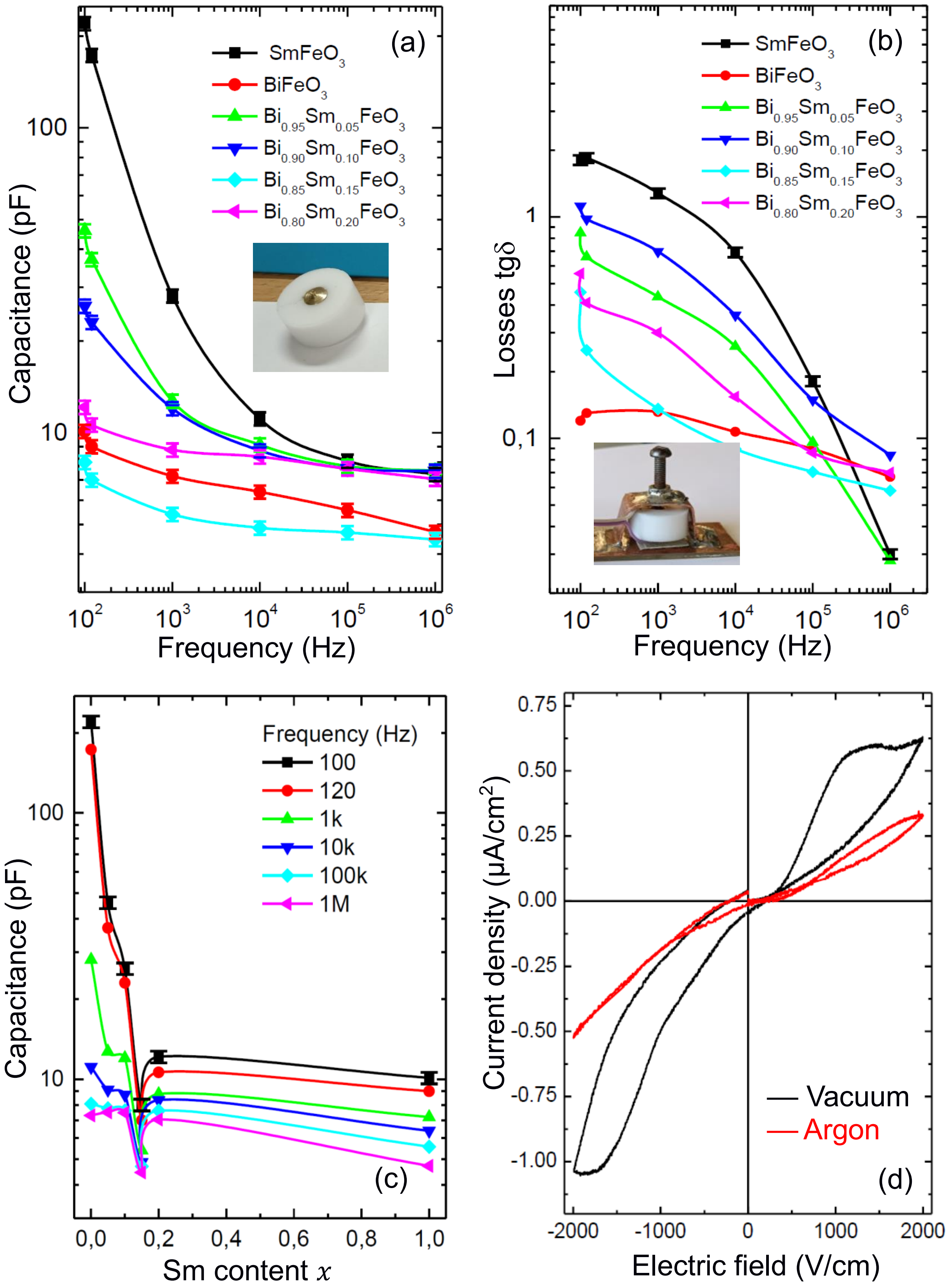

**FIGURE 13.** Frequency dependence of the effective capacitance **(a)** and dielectric losses tgδ **(b)** measured for the $Bi_{1-x}Sm_xFeO_3$ nanoparticles pressed in Teflon cells. **(c)** Dependences of the effective capacitance on the Sm content $x$ measured for different frequencies of applied bias, listed in the legend. **(d)** Typical current-

voltage curves measured after 100 hours in the high vacuum (the black curve), and after the for-vacuum pumping and 30 minutes in the Ar atmosphere (the red curve).

## IV. CONCLUSIONS

Using the FSM-LGDK-SH approach for the description of coupled polar and antipolar long-range orders in ferroics, we calculated analytically the phase diagrams and polar properties of spherical $Bi_{1-x}Sm_xFeO_3$ nanoparticles covered by surface ions in dependence on their size, density of ions, samarium content $x$ and temperature. We revealed that the size effects and ferro-ionic coupling govern the appearance and stability conditions of the long-range ordered FE, reentrant FEI and AFE phases in the nanoparticles. Also, we predict that it is possible to control the reentrant transition between the FEI and the FE phase by the size of the $Bi_{1-x}Sm_xFeO_3$ nanoparticle, and/or by the dielectric properties of environment and/or by the density of the surface ionic-electronic charge. Appeared that the influence of the surface ionic-electronic charge on the reentrant phase transitions is the strongest.

Calculated phase diagrams are in a qualitative agreement with the XRD phase analysis, EPR and FTIR spectroscopy of the $Bi_{1-x}Sm_xFeO_3$ nanopowders sintered by the solution combustion method. In particular, the $R3c$ phase, where the long-range polar and/or antipolar ordering can exist, completely disappears in the $Bi_{1-x}Sm_xFeO_3$ nanoparticles for $x > 0.15$ in accordance with XRD analysis and analytical calculations, which predict the transition to the NP phase at $x \approx 0.15$. The shape of the EPR spectra of the studied $Bi_{1-x}Sm_xFeO_3$ nanoparticles undergoes three distinctive changes at $x = 0.10$ (the line with the effective g-factor $g_{eff} \approx 6$ appears at FEI-AFE transition), at $x = 0.15$ (when the line with $g_{eff} \approx 2$ has the minimal width and height near the AFE-NP transition), and at $x > 0.2$ (when the line with $g_{eff} \approx 6$ disappears in the “deep” NP phase). It follows from the analysis of the FTIR spectra, that the complete phase transition to the NP orthorhombic phase occurs when (16 – 20)% of $Bi^{3+}$ ions are substituted with $Sm^{3+}$ ions, being in a qualitative agreement with the AFE-NP phase transition calculated at $x \approx 0.15$.

The effective capacitance of the pressed $Bi_{1-x}Sm_xFeO_3$ nanopowders, measured in ambient conditions, have a pronounced minimum at $x = 0.15$ observed for all frequences from 100 Hz to 1 MHz, and the fracture at $x = 0.1$ observed in the frequency range 100 Hz – 100 kHz, or local maximum observed at $x = 0.1$ for 1 MHz. The observed fracture or maximum at $x = 0.1$ is in a qualitative agreement with the calculated maximum of the dielectric susceptibility near $x = 0.1$, which corresponds to the poly-domain FEI-AFE transition. A possible reason for the effective capacitance nonmonotonic behavior and pronounced minimum at $x = 0.15$ (unexplained by the simplified theory presented in this work) may be the morphotropic AFE-NP phase boundary. The question requires further experimental and theoretical studies.

Typical I-V curves, measured after 100 hours in the high vacuum, demonstrate the pronounced loop opening in comparison with the I-V curves measured after the for-vacuum pumping and 30 minutes in the Ar atmosphere. This behavior indicates the significant role of the ferro-ionic coupling in the pressed $Bi_{1-x}Sm_xFeO_3$ nanopowders, being in an agreement with the experimental observations of the strong ferro-ionic coupling in other ferroelectrics in high-vacuum conditions [23].

The possibility to control the phase diagrams of the $Bi_{1-x}Sm_xFeO_3$ nanoparticles by the ferro-ionic coupling is the most interesting and promising for applications, because this allows to control the nanoparticles polar properties by the adsorption of ions from the ambient medium.

**Acknowledgements.** Authors are sincerely grateful to the Referees for the very useful discussions and suggestions. The work of A.N.M., F.O.M., O.S.P., E.V.L., A.D.Y., M.V.R., Y.O.Z. L.P.Y. and E.A.E. are funded by the National Research Foundation of Ukraine (projects "Manyfold-degenerated metastable states of spontaneous polarization in nanoferroics: theory, experiment and perspectives for digital nanoelectronics", grant N 2023.03/0132 and "Silicon-compatible ferroelectric nanocomposites for electronics and sensors", grant N 2023.03/0127). This effort (problem statement and general analysis, S.V.K.) was supported as part of the center for 3D Ferroelectric Microelectronics (3DFeM), an Energy Frontier Research Center funded by the U.S. Department of Energy (DOE), Office of Science, Basic Energy Sciences under Award Number DE-SC0021118. Numerical results presented in the work are obtained and visualized using a specialized software, Mathematica 14.0 [66].

**Authors' contribution.** A.N.M. and S.V.K. generated the research idea, formulated the problem, and wrote the manuscript draft. A.N.M. performed analytical calculations and prepared corresponding figures. E.A.E. wrote the codes. I.V.F. prepared the samples and characterized them by XRD. Y.O.Z. and L.P.Y. performed EPR measurements. O.S.P. performed electrophysical measurements. L.D. performed TEM measurements. E.V.L., M.V.R., A.D.Y. and F.O.M. performed FTIR spectroscopy. All co-authors discussed the results and worked on the manuscript improvement.

## References

[1] M. Fiebig, Revival of the magnetoelectric effect. Journal of physics D: applied physics **38**, R123 (2005), https://doi.org/10.1088/0022-3727/38/8/R01

[2] A.Y. Borisevich, E.A. Eliseev, A.N. Morozovska, C.-J. Cheng, J.-Y. Lin, Y.H. Chu, D. Kan, I. Takeuchi, V. Nagarajan, S.V. Kalinin. Atomic-scale evolution of modulated phases at the ferroelectric–antiferroelectric morphotropic phase boundary controlled by flexoelectric interaction. Nature Communications. **3**, 775 (2012), https://doi.org/10.1038/ncomms1778

[3] M. Fiebig, T. Lottermoser, D. Meier, M. Trassin. The evolution of multiferroics. Nat Rev Mater **1**, 16046 (2016), https://doi.org/10.1038/natrevmats.2016.46

[4] T. K. Lin, C. Y. Shen, C. C. Kao, C. F. Chang, H. W. Chang, C. R. Wang, and C. S. Tu. Structural evolution, ferroelectric, and nanomechanical properties of $Bi_{1-x}Sm_xFeO_3$ films (x= 0.05–0.16) on glass substrates. Journal of Alloys and Compounds, **787**, 397 (2019), https://doi.org/10.1016/j.jallcom.2019.02.008

[5] P. Sharma, A. N. Morozovska, E. A. Eliseev, Qi Zhang, D. Sando, N. Valanoor, and J. Seidel. Specific Conductivity of a Ferroelectric Domain Wall. ACS Appl. Electron. Mater. **4**, 6, 2739–2746. (2022) (https://pubs.acs.org/doi/10.1021/acsaelm.2c00261)

[6] A. N. Morozovska, E. A. Eliseev, P.S.Sankara Rama Krishnan, A. Tselev, E. Strelkov, A. Borisevich, O. V. Varenyk, N. V. Morozovsky, P. Munroe, S. V. Kalinin and V. Nagarajan. Defect thermodynamics and kinetics in thin strained ferroelectric films: the interplay of possible mechanisms. Phys.Rev.B **89**, 054102 (2014)**,** https://doi.org/10.1103/PhysRevB.89.054102

[7] P.S. Sankara Rama Krishnan, A. N. Morozovska, E. A. Eliseev, Q. M. Ramasse, D. Kepaptsoglou, W.-I. Liang, Y.-H. Chu, P. Munroe and V. Nagarajan. Misfit strain driven cation inter-diffusion across an epitaxial multiferroic thin film interface. Journal of Applied Physics, **115**, 054103 (2014), https://doi.org/10.1063/1.4862556

[8] R. Maran, S. Yasui, E. A. Eliseev, M. D. Glinchuk, A. N. Morozovska, H. Funakubo, I. Takeuchi, N. Valanoor. Interface control of a morphotropic phase boundary in epitaxial samarium-modified bismuth ferrite superlattices. Phys.Rev. **B 90**, 245131 (2014), https://doi.org/10.1103/PhysRevB.90.245131

[9] R. Maran, S. Yasui, E. Eliseev, A. Morozovska, F. Hiroshi, T.i Ichiro, V. Nagarajan. Enhancement of dielectric properties in epitaxial bismuth ferrite – bismuth samarium ferrite superlattices**.** Advanced Electronic Materials. (2016)**,** https://doi.org/10.1002/aelm.201600170

[10] J. L. Ortiz-Quiñonez, et al, Easy synthesis of high-purity $BiFeO_3$ nanoparticles: new insights derived from the structural, optical and magnetic characterization. *Inorg. Chem.* 2013, 52, 18, 10306–10317, https://doi.org/10.1021/ic400627c

[11] C. Huang, Y. Nie, R. Han, X. Yang, L. Zhuang, J. Wang, X. Xi, and J. Wan. Preparation, characterization, and mechanism for enhanced photocatalytic performance in $Bi_{1-x}Sm_xFeO_3$ nanoparticles. Journal of Sol-Gel Science and Technology, **110**, 736 (2024), https://doi.org/10.1007/s10971-024-06392-y

[12] M. Kumari, M. Chahar, S. Shankar, and O. P. Thakur. Temperature dependent dielectric, ferroelectric and energy storage properties in $Bi_{0.5}Na_{0.5}TiO_3$ (BNT) nanoparticles. Materials Today: Proceedings **67**, 688 (2022), https://doi.org/10.1016/j.matpr.2022.06.542

[13] Z. Fan, S. Gao, Y. Chang, D. Wang, X. Zhang, H. Huang, Y. He, and Q. Zhang. Ultra-superior high-temperature energy storage properties in polymer nanocomposites via rational design of core–shell structured inorganic antiferroelectric fillers. Journal of Materials Chemistry A **11**, 7227 (2023), https://doi.org/10.1039/D2TA09658G

[14] S. Chaturvedi, S. K. Singh, P. Shyam, M. M. Shirolkar, S. Krishna, R. Boomishankar, and S. Ogale. Nanoscale $LuFeO_3$: shape dependent ortho/hexa-phase constitution and nanogenerator application. Nanoscale **10**, 21406 (2018), https://doi.org/10.1039/C8NR07825D

[15] L. Baudry, I. Lukyanchuk, V. M. Vinokur. Ferroelectric symmetry-protected multibit memory cell. Scientific Reports, **7,** Article number 42196 (2017), https://doi.org/10.1038/srep42196

[16] J. Mangeri, Y. Espinal, A. Jokisaari, S. P. Alpay, S. Nakhmanson, and O. Heinonen. “Topological phase transformations and intrinsic size effects in ferroelectric nanoparticles”. Nanoscale, **9**, 1616 (2017), https://doi.org/10.1039/C6NR09111C

[17] A. N. Morozovska, E. A. Eliseev, D. Chen, C. T. Nelson, and S. V. Kalinin. Building Free Energy Functional from Atomically-Resolved Imaging: Atomic Scale Phenomena in La-doped $BiFeO_3$. Phys.Rev. B, **99**, 195440 (2019); https://link.aps.org/doi/10.1103/PhysRevB.99.195440

[18] J. Zhou, H.-H. Huang, S. Kobayashi, S. Yasui, Ke Wang, E. Eliseev, A. Morozovska, Pu Yu; I. Takeuchi, Z. Hong, D. Sando, Qi Zhang, N. Valanoor. An Emergent Quadruple Phase Ensemble in doped Bismuth Ferrite Thin Films through Site and Strain Engineering. Advanced Functional Materials, 2403410 (2024) https://doi.org/10.1002/adfm.202403410

[19] A. Raghavan, R. Pant, I. Takeuchi, E. A. Eliseev, M. Checa, A. N. Morozovska, M. Ziatdinov, S. V. Kalinin, Y. Liu, Evolution of ferroelectric properties in $Sm_xBi_{1-x}FeO_3$ via automated Piezoresponse Force Microscopy across combinatorial spread libraries, (2024), https://doi.org/10.1021/acsnano.4c06380

[20] G.B. Stephenson and M.J. Highland, Equilibrium and stability of polarization in ultrathin ferroelectric films with ionic surface compensation. Physical Review **B**, **84** (6), 064107, (2011) https://doi.org/10.1103/PhysRevB.84.064107

[21] M. J. Highland, T. T. Fister, D. D. Fong, P. H. Fuoss, Carol Thompson, J. A. Eastman, S. K. Streiffer, and G. B. Stephenson, Equilibrium polarization of ultrathin PbTiO3 with surface compensation controlled by oxygen partial pressure, Physical Review Letters,**107**, no. 18, (2011) 187602.

[22] S. M. Yang, A. N. Morozovska, R. Kumar, E. A. Eliseev, Ye Cao, L. Mazet, N. Balke, S. Jesse, R. Vasudevan, C. Dubourdieu, S. V. Kalinin. Mixed electrochemical-ferroelectric states in nanoscale ferroelectrics. Nature Physics **13**, 812 (2017), https://doi.org/10.1038/nphys4103

[23] K. P. Kelley, A. N. Morozovska, E. A. Eliseev, Y. Liu, S. S. Fields, S. T. Jaszewski, T. Mimura, J. F. Ihlefeld, S. V. Kalinin. Ferroelectricity in Hafnia Controlled via Surface Electrochemical State. Nature Materials **22**, 1144 (2023), https://doi.org/10.1038/s41563-023-01619-9

[24] A. N. Morozovska, E. A. Eliseev, N. V. Morozovsky, and S. V. Kalinin. Ferroionic states in ferroelectric thin films. Phys. Rev. B **95**, 195413 (2017), https://doi.org/10.1103/PhysRevB.95.195413

[25] A. N. Morozovska, E. A. Eliseev, A. I. Kurchak, N. V. Morozovsky, R. K. Vasudevan, M. V. Strikha, and S. V. Kalinin. Effect of surface ionic screening on polarization reversal scenario in ferroelectric thin films: crossover from ferroionic to antiferroionic states. Phys. Rev. B **96**, 245405 (2017), https://doi.org/10.1103/PhysRevB.96.245405

[26] A. N. Morozovska, E. A. Eliseev, I. S. Vorotiahin, M. V. Silibin, S. V. Kalinin and N. V. Morozovsky. Control of polarization hysteresis temperature behavior by surface screening in thin ferroelectric films. Acta Mater. **160**, 57 (2018), https://doi.org/10.1016/j.actamat.2018.08.041

[27] A. N. Morozovska, E. A. Eliseev, A. Biswas, H. V. Shevliakova, N. V. Morozovsky, and S. V. Kalinin. Chemical control of polarization in thin strained films of a multiaxial ferroelectric: phase diagrams and polarization rotation. Phys. Rev. B **105**, 094112 (2022), https://doi.org/10.1103/PhysRevB.105.094112

[28] A. N. Morozovska, S. V. Kalinin, M. E. Yelisieiev, J. Yang, M. Ahmadi, E. A. Eliseev, and Dean R. Evans. Dynamic control of ferroionic states in ferroelectric nanoparticles. Acta Materialia **237**, 118138 (2022), https://doi.org/10.1016/j.actamat.2022.118138

[29] A. N. Morozovska, E. A. Eliseev, A. Biswas, N. V. Morozovsky, and S.V. Kalinin. Effect of surface ionic screening on polarization reversal and phase diagrams in thin antiferroelectric films for information and energy storage. Phys. Rev. Appl. **16**, 044053 (2021), https://doi.org/10.1103/PhysRevApplied.16.044053

[30] A. Biswas, A. N. Morozovska, M. Ziatdinov, E. A. Eliseev and S. V. Kalinin. Multi-objective Bayesian optimization of ferroelectric materials with interfacial control for memory and energy storage applications J. Appl. Phys. **130**, 204102 (2021), https://doi.org/10.1063/5.0068903

[31] D. V. Karpinsky, E. A. Eliseev, F. Xue, M. V. Silibin, A. Franz, M. D. Glinchuk, I. O. Troyanchuk, S. A. Gavrilov, V. Gopalan, L.-Q. Chen, and A.N. Morozovska. "Thermodynamic potential and phase diagram for multiferroic bismuth ferrite ($BiFeO_3$)". npj Computational Materials **3**:20 (2017); https://doi.org/10.1038/s41524-017-0021-3

[32] See Supplemental Materials for calculation details, samples preparation and characterization [URL will be provided by Publisher]

[33] A. K. Tagantsev and G. Gerra. Interface-induced phenomena in polarization response of ferroelectric thin films. J. Appl. Phys. **100**, 051607 (2006), https://doi.org/10.1063/1.2337009

[34] J.-J. Wang, B. Wang, and L.-Q. Chen. Understanding, Predicting, and Designing Ferroelectric Domain Structures and Switching Guided by the Phase-Field Method. Ann. Rev. Mater. Res. **49**, 127 (2019).

[35] Flexoelectricity in Solids: From Theory to Applications. Ed. by A.K. Tagantsev and P.V. Yudin, World Scientific (2016)

[36] L. D. Landau, and I. M. Khalatnikov. On the anomalous absorption of sound near a second order phase transition point. Dokl. Akad. Nauk SSSR, vol. 96, pp. 469-472 (1954).

[37] R. Kretschmer and K.Binder. Surface effects on phase transitions in ferroelectrics and dipolar magnets. Phys. Rev. B **20**, 1065 (1979), https://doi.org/10.1103/PhysRevB.20.1065.

[38] C.-L. Jia, V. Nagarajan, Jia-Qing He, L. Houben, T. Zhao, R. Ramesh, K. Urban & R. Waser, Unit-cell scale mapping of ferroelectricity and tetragonality in epitaxial ultrathin ferroelectric films. Nature Materials **6**, 64 (2007), https://doi.org/10.1038/nmat1808.

[39] E. A. Eliseev, A. V. Semchenko, Y. M. Fomichov, M. D. Glinchuk, V. V. Sidsky, V. V. Kolos, Yu. M. Pleskachevsky, M. V. Silibin, N. V. Morozovsky, A. N. Morozovska. Surface and finite size effects impact on the phase diagrams, polar and dielectric properties of (Sr,Bi)$Ta_2O_9$ ferroelectric nanoparticles. J. Appl. Phys. **119**, 204104 (2016), https://doi.org/10.1063/1.4952707

[40] J. Bardeen, Surface states and rectification at a metal semi-conductor contact, Phys. Rev. **71**, 717 (1947), https://doi.org/10.1103/PhysRev.71.717

[41] Y.A. Genenko, O. Hirsch, and P. Erhart, Surface potential at a ferroelectric grain due to asymmetric screening of depolarization fields, J. Appl. Phys. **115**, 104102 (2014), https://doi.org/10.1063/1.4867984

[42] K.Y. Foo, and B. H. Hameed, Insights into the modeling of adsorption isotherm systems, Chemical Engineering Journal **156,** 2 (2010), https://doi.org/10.1016/j.cej.2009.09.013

[43] A. N. Morozovska, Y. M. Fomichov, P. Maksymovych, Yu. M. Vysochanskii, and E. A. Eliseev. Analytical description of domain morphology and phase diagrams of ferroelectric nanoparticles. Acta Materialia **160**, 109-120 (2018) https://doi.org/10.1016/j.actamat.2018.08.051

[44] I. S. Vorotiahin, E. A. Eliseev, Q. Li, S. V. Kalinin, Y. A. Genenko and A. N. Morozovska. Tuning the Polar States of Ferroelectric Films via Surface Charges and Flexoelectricity. Acta Materialia **137** (15), 85–92 (2017) https://doi.org/10.1016/j.actamat.2017.07.033

[45] E. A. Eliseev, S. V. Kalinin, A. N. Morozovska. Ferro-ionic States and Domains Morphology in $Hf_xZr_{1-x}O_2$ Nanoparticles (2024), https://doi.org/10.48550/arXiv.2410.04476

[46] Z. Hu, D. Chen, S. Wang, N. Zhang, L. Qin, and Y. Huang. Facile synthesis of Sm-doped BiFeO3 nanoparticles for enhanced visible light photocatalytic performance. Materials Science and Engineering: B **220**, 1-12 (2017), https://doi.org/10.1016/j.mseb.2017.03.005

[47] K. Papadopoulos, E. Myrovali, A. Dubey, L. Malletzidou, D.C. Lupascu, V.V. Shvartsman, U. Wiedwald, and M. Angelakeris. Control of physical properties in BiFeO3 nanoparticles via Sm3+ and Co2+ ion doping. Nanotechnology **35**, 015707 (2023), https://doi.org/10.1088/1361-6528/acfcc2

[48] K. Papadopoulos, E. Myrovali, L. Malletzidou, D. Karfaridis, I. Tarasov, G. Vourlias, C. Sarafidis, M. Spasova, M. Farle, U. Wiedwald, M. Angelakeris. Control of multiferroic features in BiFeO3 nanoparticles by facile synthetic parameters. Ceramics International **49**, 18552-18564 (2023). https://doi.org/10.1016/j.ceramint.2023.02.229

[49] M. Kumar, S. Chauhan, and H. Pandey. Effect of $Gd^{3+}$ substitution on structural, morphological, and magnetic properties of BiFeO3 nanoparticles. Journal of Sol-Gel Science and Technology **109**, 272 (2024), https://doi.org/10.1007/s10971-023-06269-6

[50] M. Kumar, and H. Pandey. Structural Phase Transformation, Magnetic and Optical Properties of $Ho^{3+}$ Substituted $BiFeO_3$ Nanoparticles. Journal of Superconductivity and Novel Magnetism **36**, 1269-1276 (2023), https://doi.org/10.1007/s10948-023-06568-7

[51] D. Maurya, *et al*, Magnetic studies of multiferroic Bi1−xSmxFeO3 ceramics synthesized by mechanical activation assisted processes, J. Phys.: Condens. Matter, **21**, 026007 (2009), https://doi.org/10.1088/0953-8984/21/2/026007

[52] M. T. Kebede *et al*, Crystal structure refinement and magnetic properties of $Sm^{3+}$ doped $BiFeO_3$ nanoparticles, Physica B: Condensed Matter, **624**, 413374, (2022), https://doi.org/10.1016/j.physb.2021.413374

[53] J. W. Lin *et al*, Electron spin resonance probed suppressing of the cycloidal spin structure in doped bismuth ferrites, Appl. Phys. Lett. **96**, 232507 (2010), https://doi.org/10.1063/1.3451463

[54] T. Castner Jr., G.S. Newell, W.C. Holton, C.P. Slichter, Note on the Paramagnetic Resonance of Iron in Glass, J. Chem. Phys. **32**, 668 (1960), https://doi.org/10.1063/1.1730779

[55] C. Ederer and N. A. Spaldin. Weak ferromagnetism and magnetoelectric coupling in bismuth ferrite. Phys. Rev. B **71**, 060401(R) (2005), https://doi.org/10.1103/PhysRevB.71.060401

[56] J.-G. Park, M. Duc Le, J. Jeong, and S. Lee. Structure and spin dynamics of multiferroic $BiFeO_3$. J. Phys.: Condens. Matter, **26**, 433202 (2014). https://doi.org/10.1088/0953-8984/26/43/433202

[57] P. Meisenheimer, G. Moore, S. Zhou, H. Zhang, X. Huang, S. Husain, X. Chen, L. W. Martin, K. A. Persson, S. Griffin, L. Caretta. Switching the spin cycloid in $BiFeO_3$ with an electric field. Nat Commun. **15**, 2903 (2024). https://doi.org/10.1038/s41467-024-47232-5

[58] B. Ruette, S. Zvyagin, A. P. Pyatakov, A. Bush, J. F. Li, V. I. Belotelov, A. K. Zvezdin, and D. Viehland. Magnetic-field-induced phase transition in $BiFeO_3$ observed by high-field electron spin resonance: Cycloidal to homogeneous spin order. Phys. Rev. B **69**, 064114 (2004), https://doi.org/10.1103/PhysRevB.69.064114

[59] C. Chen, J. Cheng, S. Yu, L. Che, Z. Meng. Hydrothermal synthesis of perovskite bismuth ferrite crystallites. J. Cryst. Growth, **291**, 135–139, (2006), https://doi.org/10.1016/j.jcrysgro.2006.02.048

[60] K. Nakamoto. Infrared and Raman spectra of inorganic and coordination compounds, Wiley-Interscience (2009), https://doi.org/10.1002/9780470405840

[61] M. Gowrishankar, D. Rajan Babu, S. Madeswaran. Effect of Gd–Ti co-substitution on structural, magnetic and electrical properties of multiferroic $BiFeO_3$. J. Magn. Magn. Mater., **418**, 54-61, (2016). https://doi.org/10.1016/j.jmmm.2016.03.085

[62] V. M. Gaikwad, S. A. Acharya. Investigation of spin phonon coupling in $BiFeO_3$ based system by Fourier transform infrared spectroscopy**.** J. Appl. Phys., **114**, 193901, (2013), https://doi.org/10.1063/1.4831676

[63] C.H. Yang, D. Kan, I. Takeuchi, V. Nagarajan, J. Seidel. Doping $BiFeO_3$: approaches and enhanced functionality. Phys. Chem. Chem. Phys., **14**, 15953, (2012). https://doi.org/10.1039/C2CP43082G

[64] S.P. Tandon and P.C. Mehta. The Infrared Absorption Spectra of Some Samarium Diketoester Complexes. Z. Naturforsch. B, **25b**, 139 - 141 (1970), https://doi.org/10.1515/znb-1970-0202

[65] K.W. Wagner, Arch Elektrotech **2**, 371 (1914); https://doi.org/10.1007/BF01657322

[66] https://www.wolfram.com/mathematica

**SUPPLEMENTAL MATERIALS to**

# "Reentrant polar phase induced by the ferro-ionic coupling in $Bi_{1-x}Sm_xFeO_3$ nanoparticles"

Anna N. Morozovska[1*], Eugene A. Eliseev[2], Igor V. Fesych[3], Yuriy O. Zagorodniy[2], Oleksandr S. Pylypchuk[1], Evgenii V. Leonenko[1], Maxim V. Rallev[1], Andrii D. Yaremkevych[1], Lesya P. Yurchenko[2], Lesya Demchenko[4], Sergei V. Kalinin[5†], and Olena M. Fesenko[1‡]

[1] Institute of Physics, National Academy of Sciences of Ukraine,46, pr. Nauky, 03028 Kyiv, Ukraine

[2]Frantsevich Institute for Problems in Materials Science, National Academy of Sciences of Ukraine, 3, str. Omeliana Pritsaka, 03142 Kyiv, Ukraine

[3]Taras Shevchenko National University of Kyiv, 01030 Kyiv, Ukraine

[4]Stockholm University, Department of Materials and Environmental Chemistry, Sweden and Ye. O. Paton Institute of Materials Science and Welding, National Technical University of Ukraine "Igor Sikorsky Kyiv Polytechnic Institute"

[5]Department of Materials Science and Engineering, University of Tennessee, Knoxville, TN, 37996, USA

## Appendix A

### A1. Calculation details

The 2-4-Landau free energy dependence on the polarization, anti-polarization and electric field is

$$F_R = \frac{1}{2}\alpha_R P^2 + \frac{1}{2}\eta A^2 + \frac{1}{4}\beta_P P^4 + \frac{1}{4}\beta_A A^4 + \frac{1}{2}\xi P^2 A^2 - PE. \qquad \text{(A.1)}$$

Equations of state could be obtained from the minimization of the free energy (A.1):

$$\alpha_R P + \beta_P P^3 + \xi P A^2 = E, \qquad \text{(A.2a)}$$

$$\eta A + \beta_A A^3 + \xi P^2 A = 0. \qquad \text{(A.2b)}$$

Dielectric susceptibility is given by expression:

$$\chi = \frac{\eta + 3\beta_A A^2 + \xi P^2}{(\alpha_R + 3\beta_P P^2 + \xi A^2)(\eta + 3\beta_A A^2 + \xi P^2) - 4\xi^2 P^2 A^2} \qquad \text{(A.3)}$$

---

* corresponding author, e-mail: anna.n.morozovska@gmail.com
† corresponding author, e-mail: sergei2@utk.edu
‡ corresponding author, e-mail: fesenko.olena@gmail.com

The spontaneous order parameters, free energy densities, stability conditions of the spatially homogeneous phases and corresponding critical field(s) in the single-domain phases of the 2-4-Landau model are listed in **Table A1.**

**Table A1.** The spontaneous order parameters, stability conditions, dielectric susceptibility, free energy densities and critical field values in different phases *.

| Phase | Paraelectric (NP) | Ferroelectric- (FE) | Ferrielectric (FEI) | Antipolar (AFE) |
|---|---|---|---|---|
| Order parameters | $P = A = 0$ | $A = 0,$ $P = \pm\sqrt{\frac{-\alpha_R}{\beta_P}}$ | $A = \pm\sqrt{-\frac{\eta\beta_P-\xi\alpha_R}{\beta_P\beta_A-\xi^2}}$ $P = \pm\sqrt{-\frac{\alpha_R\beta_A-\xi\eta}{\beta_P\beta_A-\xi^2}}$ | $A = \pm\sqrt{-\frac{\eta}{\beta_A}},$ $P = 0.$ |
| Stability conditions | $\alpha_R > 0,$ $\eta > 0$ | $\alpha_R < 0,$ $\eta - \xi\frac{\alpha_R}{\beta_P} > 0$ | $\eta\beta_P - \xi\alpha_R < 0,$ $\alpha_R\beta_A - \xi\eta < 0,$ $\beta_P\beta_A > \xi^2$ | $\eta < 0,$ $\alpha_R\beta_A - \eta\xi > 0$ $\xi > \sqrt{\beta_P\beta_A}$ |
| Free energy | $f_D = 0$ | $f_P = -\frac{{\alpha_R}^2}{4\beta_P}$ | $f_{PA} = \frac{-\beta_P{\alpha_R}^2-\beta_A\eta^2+\xi\alpha_R\eta}{4(\beta_P\beta_A-\xi^2)}$ | $f_A = -\frac{\eta^2}{4\beta_A}$ |
| Susceptibility | $\chi = \frac{1}{\alpha_R}$ | $\chi = \frac{1}{-2\alpha_R}$ | $\chi = \frac{1}{-2\left(\alpha_R-\frac{\xi\,\eta}{\beta_A}\right)}$ | $\chi = \frac{1}{\alpha_R-\frac{\xi\,\eta}{\beta_A}}$ |
| Coercive or critical field(s) | $E_c = 0$ | $E_c = \frac{\pm 2}{3\sqrt{3}}\frac{(-\alpha_R)^{3/2}}{\beta_P}$ | $E_c = \frac{\pm 2}{3\sqrt{3}}\frac{\left(-\alpha_R+\frac{\xi}{\beta_A}\eta\right)^{3/2}}{\beta_P-\frac{\xi^2}{\beta_A}}$ | $E_{c1} = \pm\left(\alpha_R - \frac{\eta}{\xi}\beta_P\right)\sqrt{-\frac{\eta}{\xi}}$ $E_{c2} = \frac{\pm 2}{3\sqrt{3}}\frac{\left(\alpha_R-\frac{\xi\,\eta}{\beta_A}\right)^{3/2}}{\frac{\xi^2}{\beta_A}-\beta_P}$ |

* For the bulk sample the coefficient $\alpha_R \to \alpha$.

The equivalence of sublattices can lead to the assumption $\beta_P = \beta_A = \beta = 1$ in Eq.(A.1) without any loss of generality. Next, the dimensionless variables can be introduced as:

$$\tilde{F}_{bulk} \equiv \frac{F_{bulk}}{\alpha_0 P_0^2} = \frac{\alpha(T,x)}{2\alpha_0}\tilde{P}^2 + \frac{1}{2}\frac{\eta(T,x)}{\alpha_0}\tilde{A}^2 + \frac{1}{4}\frac{\beta P_0^4}{\alpha_0 P_0^2}\left(\tilde{P}^4 + \tilde{A}^4\right) + \frac{\xi\,P_0^2}{2\alpha_0}\tilde{P}^2\tilde{A}^2 - \frac{\vec{P}\vec{E}}{\alpha_0 P_0^2} = \frac{1}{2}\tilde{\alpha}(T,x)\tilde{P}^2 + \frac{1}{2}\tilde{\eta}(T,x)\tilde{A}^2 + \frac{1}{4}\left(\tilde{P}^4 + \tilde{A}^4\right) + \frac{1}{2}\tilde{\xi}\tilde{P}^2\tilde{A}^2 - \tilde{\vec{P}}\tilde{\vec{E}}, \tag{A.4a}$$

$$\tilde{\alpha}(T,x) = \frac{\alpha(T,x)}{\alpha_0},\ \ \tilde{\eta}(T,x) = \frac{\eta(T,x)}{\alpha_0},\ \ \tilde{\xi} = \frac{\xi}{\beta},\ \ P_0^2 = \frac{\alpha_0}{\beta},\ \ \tilde{P}^2 = \frac{P^2}{P_0^2},\ \ \tilde{A}^2 = \frac{A^2}{P_0^2},\ \ \tilde{E} = \frac{E}{\alpha_0 P_0}. \tag{A.4b}$$

$$\frac{\alpha_R(T,x,R)}{\alpha_0} = \frac{T}{T_C} - \exp\left[-\left(\frac{x}{x_C}\right)^4\right] + \frac{1}{\varepsilon_0\alpha_0}\frac{1}{\{\varepsilon_b+2\varepsilon_M+\varepsilon_s(R/\lambda)\}}. \tag{A.4c}$$

## A2. Field dependences of the polar and antipolar order parameters

The quasi-static dependences, $P_3(E_3)$, calculated for different Sm content $x$ in the $Bi_{1-x}Sm_xFeO_3$ nanoparticles, are shown in **Fig. A1** for the case of the relatively small ($\xi \ll 1$) and big ($\xi \gg 1$) coupling constant $\xi$.

It is seen from **Figs. A1(a)-(b)**, calculated for $\xi \ll 1$, that the decrease of the $n_i$ (from $10^{19}$m$^{-2}$ to $6 \cdot 10^{17}$m$^{-2}$) leads to the ferroelectricity disappearance and to the appearance of the paraelectric-like response of polarization to the external field.

It is seen from **Fig. A1(c)-(d)**, calculated for $\xi \gg 1$, the quasi-rectangular ferroelectric single loops at first transforms to the antiferroelectric-type double-loops and then to the linear hysteresis-less curves with the decrease of $n_i$ from $10^{19}$m$^{-2}$ to $6 \cdot 10^{17}$m$^{-2}$. Further decrease of $n_i$ leads to the gradual shrinking of the double loops, which eventually transform in the linear dielectric-like polarization response to the external field for $n_i \leq 10^{17}$m$^{-2}$.

It is seen from **Figs. A2**, calculated for $\xi \ll 1$ and relatively high density of surface ions ($n_i = 2 \cdot 10^{18}$m$^{-2}$), that the increase of Sm content $x$ from 0 to 0.10 leads to the gradual transition of the quasi-rectangular ferroelectric-type single loops of the polarization to the hysteresis-less paraelectric curves.

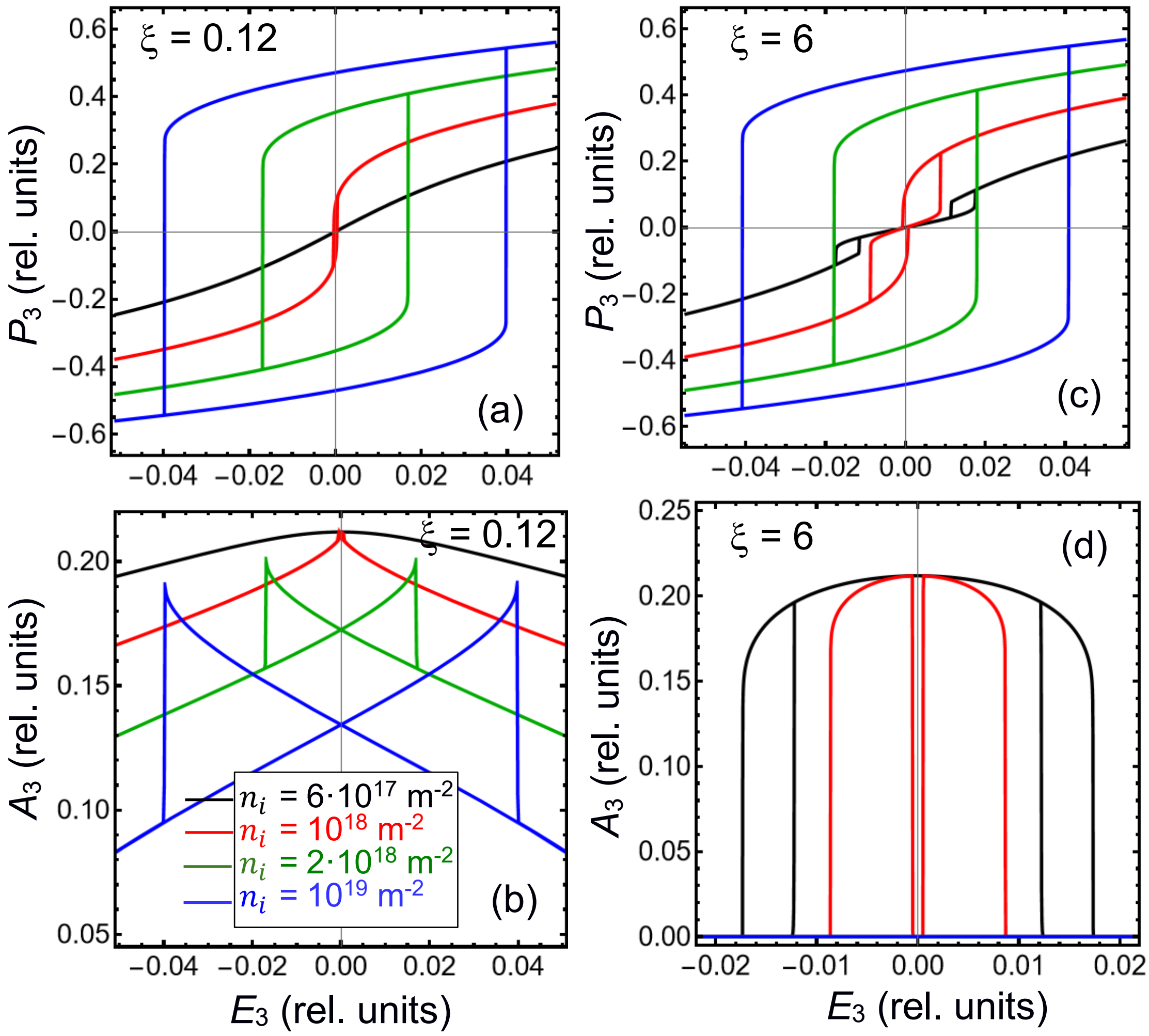

**FIGURE A1**. The quasi-equilibrium dependences of the polarization $P_3$ **(a, c)** and anti-polarization $A_3$ **(b, d)** on the external field $E$ calculated for the $Sm_xBi_{1-x}FeO_3$ nanoparticles, with Sm content $x = 0.09$ and for $n_i = 6 \cdot 10^{17}$m$^{-2}$ (the black curves), $10^{18}$m$^{-2}$ (the red curves), $2 \cdot 10^{18}$m$^{-2}$ (the green curves) and $10^{19}$m$^{-2}$ (the blue curves). The FE-AFE coupling constant $\xi = 0.121$ for plots **(a, b)** and $\xi = 6$ for plots **(c, d)**, while the other parameters are $\varepsilon_M = 30$, $R = 5$ nm and $T$ =25°C. Other parameters are the same as in **Fig. 3** in the main text.

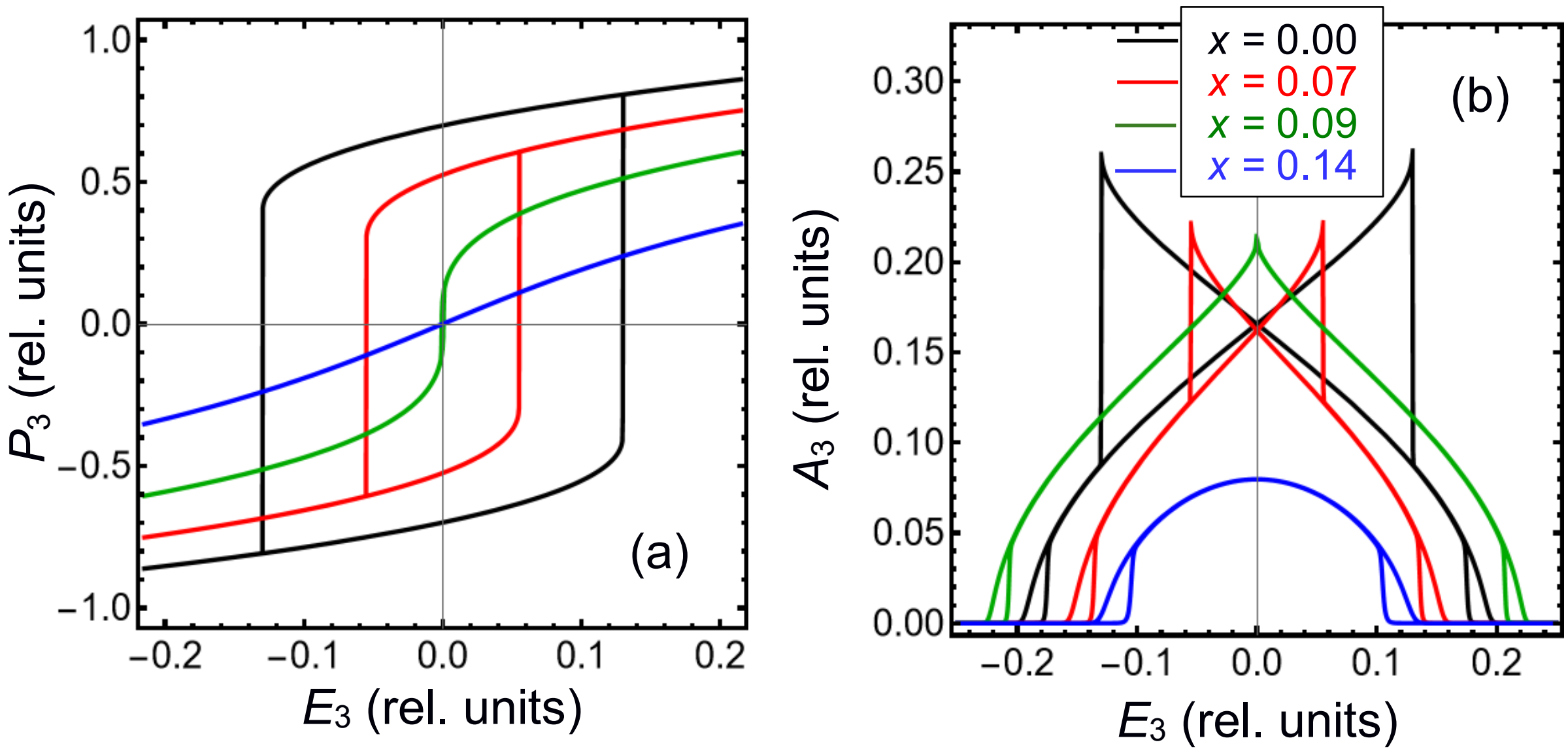

**FIGURE A2**. The quasi-equilibrium dependences of the polarization $P_3$ **(a)** and anti-polarization $A_3$ **(b)** in the external field $E$ calculated for the $Sm_xBi_{1-x}FeO_3$ nanoparticles, which Sm content $x = 0$ (the black curves), 0.07 (the red curves), 0.09 (the green curves) and 0.14 (the blue curves). The FE-AFE coupling constant $\xi = 0.121$, $\varepsilon_M = 30$, $R = 5$ nm, $T$ =25°C and $n_i = 10^{18}$m$^{-2}$. Other parameters are the same as in **Fig. 3** in the main text.

Considering that a given domain structure is characterized by polarization contrast and amplitude, period and/or other features of domain morphology, the diagram, illustrating the domain structure changes with the Sm content $x$, can be calculated by the FEM. In the case one should fix the particle size $R$, temperature $T$, ambient dielectric constant $\varepsilon_M$ and change the parameters $n_i$ and $x$. Each pair $\{n_i, x\}$ corresponds to a concrete domain pattern, paraelectric or single-domain state. **Figure A3** shows an example of such patterns calculated for the $T$ =25°C, $R$ = 20 nm and $\varepsilon_M$ =30. It is seen that the single-domain FE phase is in the top right corner of the figure. The poly-domain (or bi-domain) states, which are stable at lower $n_i$ and higher $x$, separate the single-domain phase from the paraelectric states. The poly-domain and bi-domain stripes gradually lose their contrast, become finer and shrink to the central part of the nanoparticle with an increase in x and/or decrease in $n_i$. The poly-domain states eventually transform into the paraelectric states, which occupy the bottom left corner of the figure.

We also notified that the view of the patterns, as well as the area of the poly-domain, single-domain and paraelectric phase, significantly depends on the $T$, $R$and $\varepsilon_M$. **Figure A3** is an example. To study the question on a quantitative level it is necessary to perform a rather time-consuming multi-parametric study of the solutions of the Euler-Lagrange problem with boundary conditions (by using FEM and ML methods).

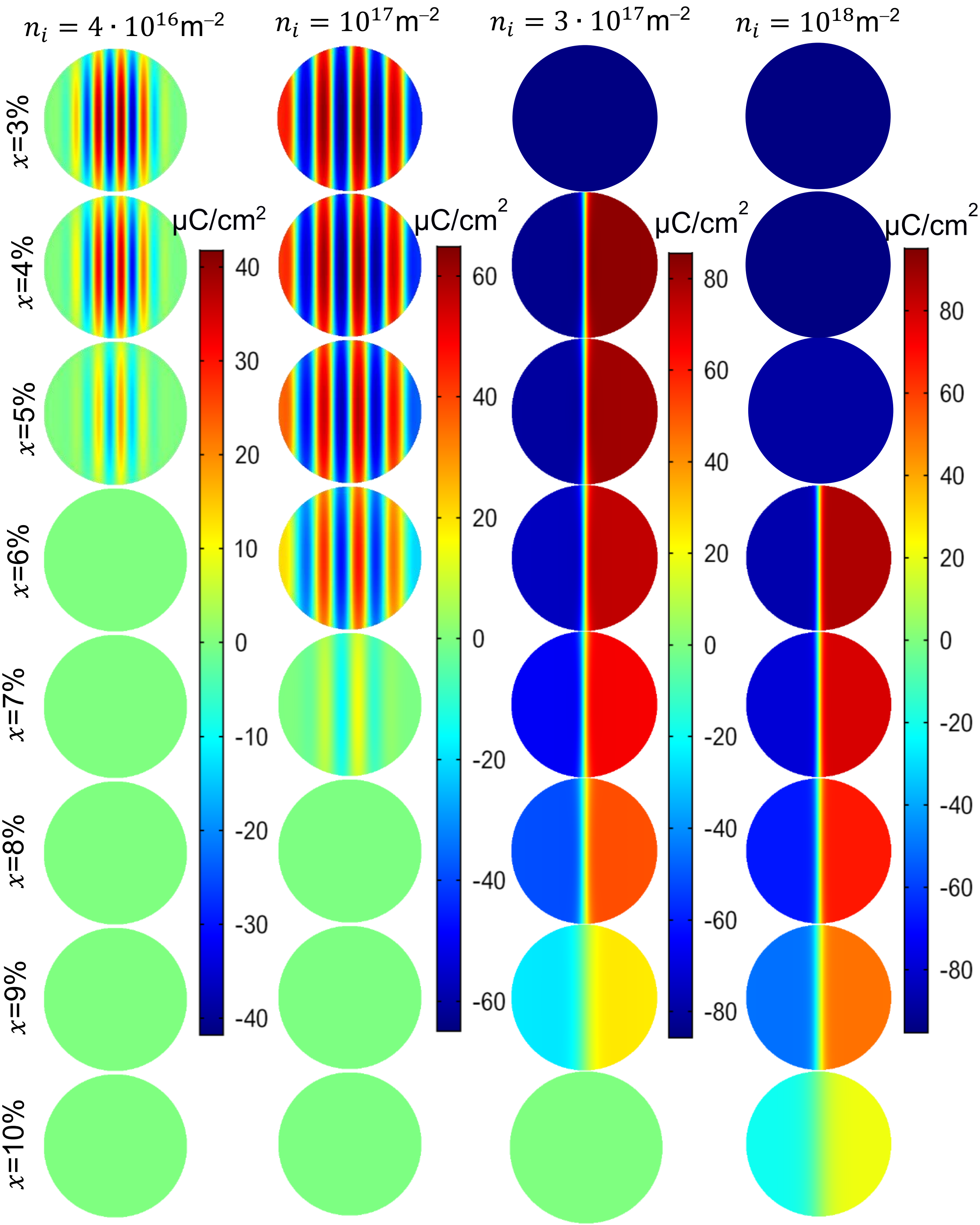

**FIGURE A3.** The distribution of polarization calculated by the FEM in the cross-section of the $Bi_{1-x}Sm_xFeO_3$ nanoparticle with different Sm content $x$ =3, 4, 5, 6, 7, 8, 9 and 10 % (different rows from the top to the bottom). The nanoparticle is covered with the ionic-electronic screening charges with the surface

density $n_i = 4 \cdot 10^{16}\text{m}^{-2}$, $10^{17}\text{m}^{-2}$ $3 \cdot 10^{17}\text{m}^{-2}$, and $10^{18}\text{m}^{-2}$ (from the left to the right column). Parameters $T$ =25°C, $R = 20$ nm and $\varepsilon_M$ =30.

Note that the features related to the domain structure remain visible in the presence of small external field and smear with the field increase (compare different curves in as **Fig. A4**).

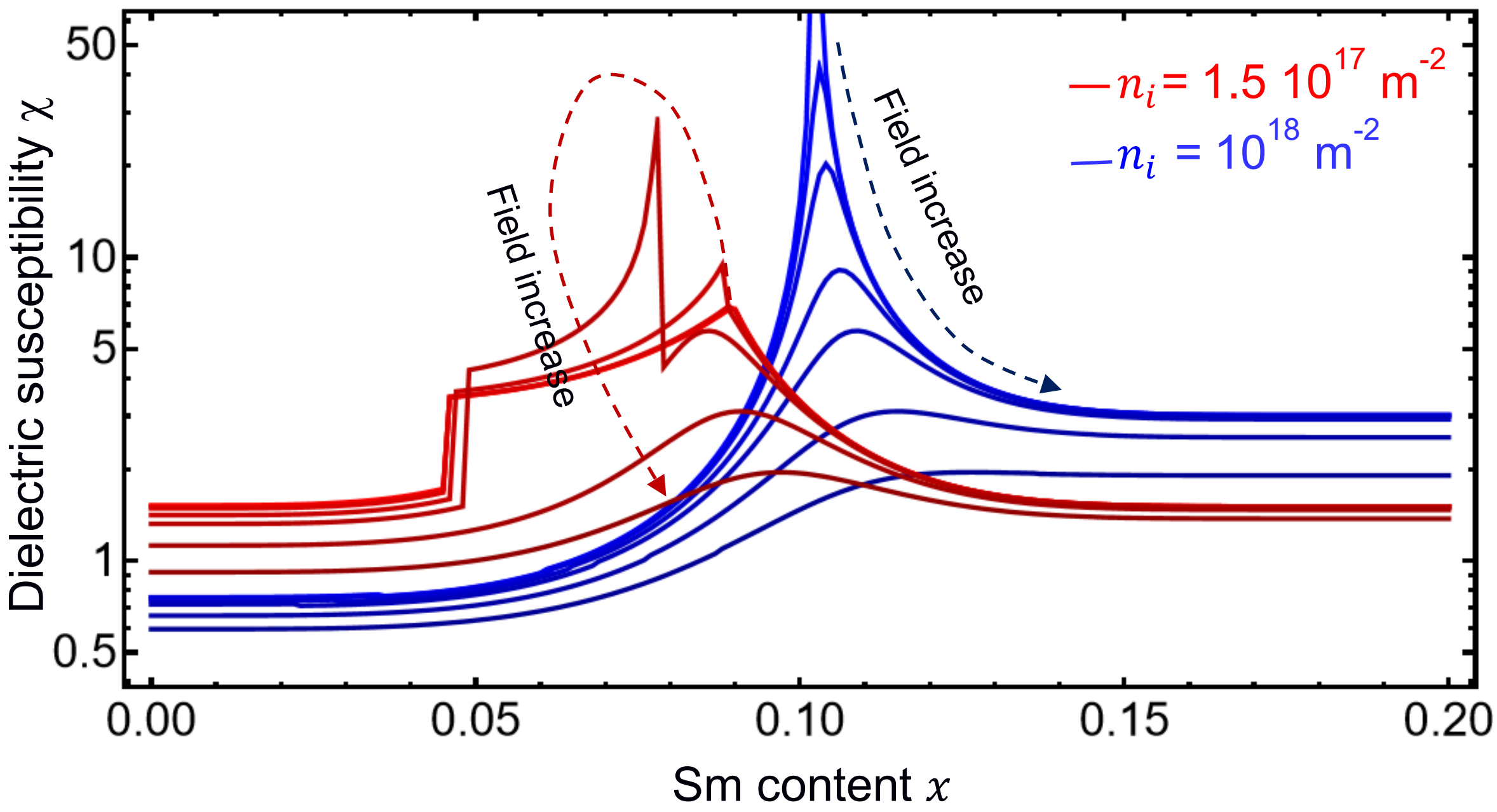

**FIGURE A4.** The dependence of the nanoparticle dielectric susceptibility on the Sm content $x$ calculated at $E = 0$ for $n_i = 1.5 \cdot 10^{17}\text{m}^{-2}$ (the red, darked red and dark-red curves) and $10^{18}\text{m}^{-2}$ (the blue, darker blue and dark-blue curves). The value of electric field $E$ = 0, 0.001,0.003, 0.01, 0.02, 0.03, 0.05 and 0.1 (in relative units). The field increases in the direction of dashed arrows. Parameters $T$ =25°C, $R$ = 20 nm and $\varepsilon_M$ =30; other parameters are the same as in **Fig. 3.**

## Appendix B. Samples preparation and characterization

A series of nanopowder samples of $Bi_{1-x}Sm_xFeO_3$ ($0 \leq x \leq 0.2$) were synthesized using the solution combustion method. Previously prepared solutions of bismuth, samarium, and iron nitrates and citric acid were used as starting components. The bismuth (III) oxide $Bi_2O_3$ (99.9 % trace metals basis, Sigma-Aldrich) and samarium (III) oxide $Sm_2O_3$ (99.9 % trace metals basis, Sigma-Aldrich) were dissolved in concentrated nitric acid $HNO_3$ (65 %, Lachema) and a small amount of deionized water was added to obtain the corresponding bismuth and samarium solutions. The iron nitrate solution was prepared by dissolving iron (III) nitrate nonahydrate $Fe(NO_3)_3 \cdot 9H_2O$ crystals in deionized water. The molar concentration of the samarium nitrate solution was determined by direct complexometric titration with Trilon B (sodium ethylenediaminetetraacetate, molarity 0.05 mol·L$^{-1}$) as a titrant and xylenol orange as an indicator. A solution containing bismuth $Bi^{3+}$ and iron $Fe^{3+}$ cations is back titrated with standard $ZnCl_2$ solution (molarity 0.048 mol·L$^{-1}$) and xylenol orange as a metal ion indicator.

First, citric acid was added to the solutions of metal nitrates which were taken in stoichiometric amounts. In this case, the number of moles of citric acid and the total number of moles of metal ions were equal. The resulting mixed solution was placed on a magnetic stirrer, stirred at a constant speed of 500 rpm, and heated to 90 °C for 2 hours. After some time, the water was evaporated and the solution turned into a dark brown gel. After further increasing the temperature to 200 °C, the gel was dried and the exothermic and self-sustained thermally induced redox reaction of xerogel combustion was observed. In the dried gel, metal nitrates provide the oxidizing agent and organic acid the reducing agent and fuel [1]. Finally, the orange powder was calcined at a temperature of 750 °C for 5 hours with a heating rate of 5°C·min$^{-1}$ and the obtained six samples were named as the BFO, SFO, BSFO-005, BSFO-010, BSFO-015 and BSFO-020, respectively (see **Table B1** and **Fig. 10(a)** in the main text).

The size of the coherent scattering regions (CSR), determined from the broadening of the XRD peaks in $Bi_{1-x}Sm_xFeO_3$ samples, shown in **Fig. B1**, indicates that the CSR size is approximately 103 nm for $x = 0$, it decreases to 55 nm for $x = 0.10$ and decreases further up to 46 nm for $x = 1$. Since the sizes of the CSR are the same order as the period of the magnetic cycloid in the pure $BiFeO_3$ (about 60 nm), it is reasonable to conclude that the size effect plays the key role in the formation of magnetoelectric adhesive containing $Bi_{1-x}Sm_xFeO_3$ nanoparticles.

**Table B1.** Samples description and characterization by XRD

| Sample acronym | Phase composition | Symmetry | Lattice parameters | CSR size |
|---|---|---|---|---|
| **BFO** | $BiFeO_3$ (72.5±1.0%) | rhombohedral (R) *R3c* (No. 161) | $a$=$b$= 5.5751(1) Å; $c$ = 13.8601(4) Å; $\alpha = \beta$ = 90°, $\gamma$ = 120 373.075(13) Å$^3$ | 103 nm |
| | $Bi_2Fe_4O_9$ (16.3±0.5%) | orthorhombic (O) *Pbam* (No. 55) | $a$ = 7.9677(7) Å; $b$ = 8.4415(6) Å; $c$ = 6.0035(5) Å; $\alpha = \beta = \gamma$ = 90° 403.798(62) Å$^3$ | |
| | $Bi_{25}FeO_{40}$ (11.2±0.6%) | cubic *I23* (No. 197) | $a$ = 10.1703(3) Å; $\alpha = \beta = \gamma$ = 90° 1051.981(61) Å$^3$ | |
| **BSFO-005** | $Bi_{0.95}Sm_{0.05}FeO_3$ (96.7±0.9%) | rhombohedral (R) *R3c* (No. 161) | $a$ = 5.5715(1) Å; $c$ = 13.8316(2) Å; $\alpha = \beta$ = 90°, $\gamma$ = 120° 371.846(10) Å$^3$ | 62 nm |
| | $Bi_2Fe_4O_9$<br>(2.2±0.3%) | orthorhombic (O)<br>*Pbam* (No. 55) | *a* = 7.9708(31) Å;<br>*b* = 8.4412(28) Å;<br>*c* = 6.0004(25) Å;<br>$\alpha = \beta = \gamma = 90°$<br>403.737(271) $Å^3$ | |
| | $Bi_{25}FeO_{40}$<br>(1.1±0.2%) | Cubic<br>*I23* (No. 197) | *a* = 10.1653(9) Å;<br>$\alpha = \beta = \gamma = 90°$<br>1050.437(173) $Å^3$ | |
| **BSFO-010** | $Bi_{0.90}Sm_{0.10}FeO_3$<br>(98.8±1.1%) | rhombohedral (R)<br>*R3c* (No. 161) | *a* = 5.5667(1) Å;<br>*c* = 13.7957(5) Å;<br>$\alpha = \beta = 90°$, $\gamma = 120°$<br>370.232(17) $Å^3$ | 55 nm |
| | $Bi_{25}FeO_{40}$<br>(1.2±0.1%) | cubic<br>*I23* (No. 197) | *a* = 10.1646(24) Å;<br>$\alpha = \beta = \gamma = 90°$<br>1050.217(436) $Å^3$ | |
| **BSFO-015** | $Bi_{0.85}Sm_{0.15}FeO_3$<br>(7.2±2%) | rhombohedral (R)<br>*R3c* (No. 161) | *a* = 5.5667(1) Å;<br>*c* = 13.7957(5) Å;<br>$\alpha = \beta = 90°$, $\gamma = 120°$<br>370.232(17) $Å^3$ | |
| | $Bi_{0.85}Sm_{0.15}FeO_3$<br>(91.8±2%) | orthorhombic (O)<br>*Pbnm* (No. 62)* | *a* = 5.4386(3) Å;<br>*b* = 5.6127(3) Å;<br>*c* = 7.8080(4) Å;<br>$\alpha = \beta = \gamma = 90°$<br>238.346(23) Å | 51 nm |
| | $Bi_{25}FeO_{40}$<br>(1.2±0.1%) | cubic (C)<br>*I23* (No. 197) | *a* = 10.1646(24) Å;<br>$\alpha = \beta = \gamma = 90°$<br>1050.217(436) $Å^3$ | |
| **BSFO-020** | $Bi_{0.8}Sm_{0.2}FeO_3$<br>(98.8±1.8%) | orthorhombic (O)<br>*Pbnm* (No. 62) | *a* = 5.4386(3) Å;<br>*b* = 5.6127(3) Å;<br>*c* = 7.8080(4) Å;<br>$\alpha = \beta = \gamma = 90°$<br>238.346(23) $Å^3$ | 50 nm |
| | $Bi_{25}FeO_{40}$<br>(1.2±0.1%) | cubic (C)<br>I23 (No. 197) | *a* = 10.1319(31) Å;<br>$\alpha = \beta = \gamma = 90°$<br>1040.115(552) $Å^3$ | |
| **SFO** | $SmFeO_3$ (100%) | orthorhombic (O)<br>*Pbnm* (No. 62) | *a* = 5.3981(2) Å,<br>*b* = 5.5928(2) Å,<br>*c* = 7.7083(3) Å,<br>$\alpha = \beta = \gamma = 90°$<br>232.719(19) $Å^3$ | 46 nm |

**Pbnm* differs from *Pnma* by the choice of coordinate setting

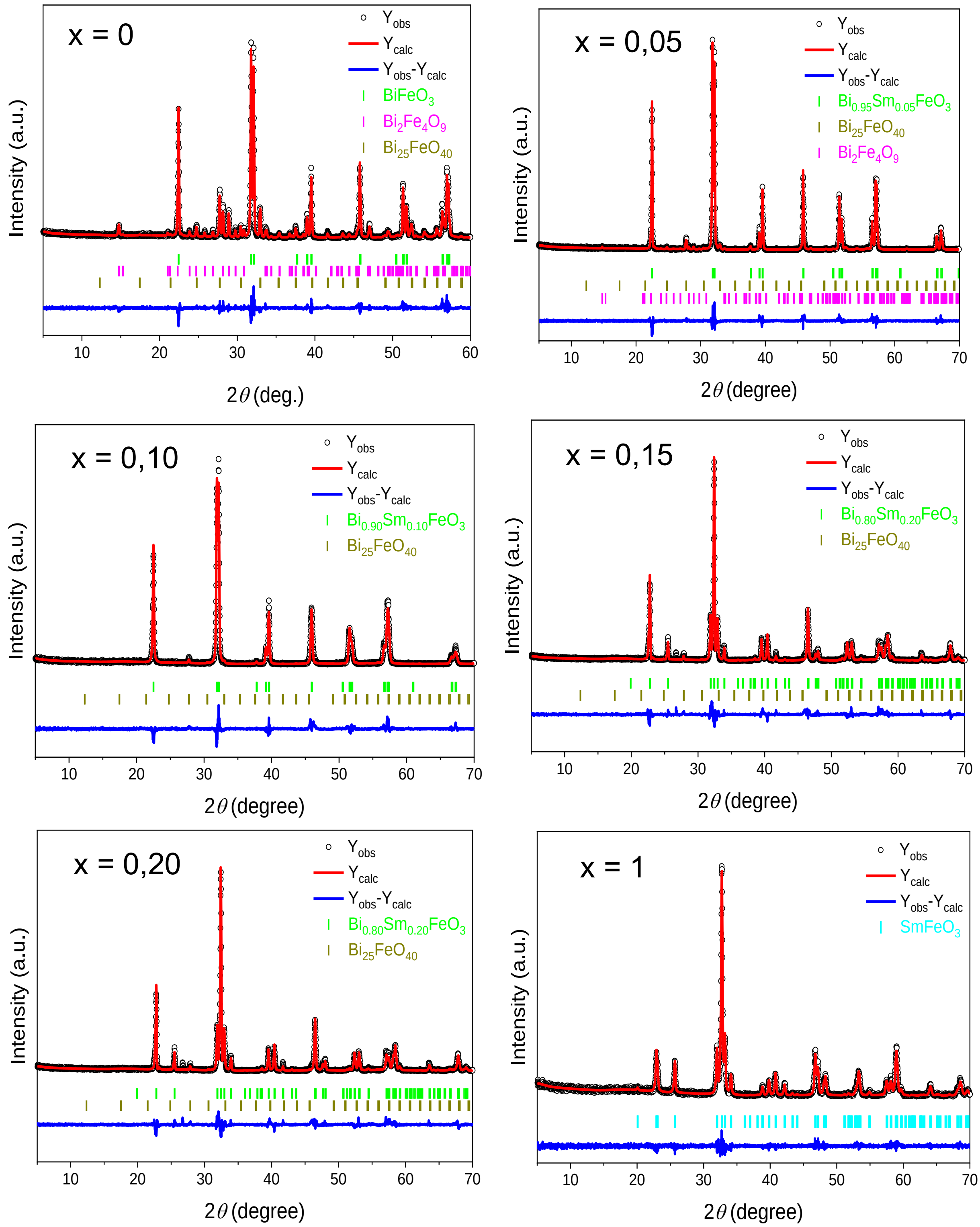

**Figure B1.** The X-ray diffraction spectra for powders with $Bi_{1-x}Sm_xFeO_3$ nanoparticles for different samarium content "x" indicated in the spectra.

Electron paramagnetic resonance (EPR) measurements were carried out on X-band Bruker Elexsys E580 spectrometer operating at 9.8 GHz frequency at room temperature. The EPR spectra of the as-prepared samples of $Bi_{1-x}Sm_xFeO_3$ nanoparticles with increasing Sm content ($x = 0, 0.1$, 0.15, 0.2 and 1) and the lines parameters are presented in **Fig. 11** and in **Table I** in the main text.

According to [2], the spin Hamiltonian of isolated $Fe^{3+}$ ion can be written as:

$$H = g\beta\vec{H}\cdot\vec{S} + D\left\{S_z^2 - \left[\frac{S(S+1)}{3}\right]\right\} + E\left(S_x^2 - S_y^2\right), \quad (B.1)$$

where $H$ is the applied magnetic field, β is Bohr magneton, S is the effective spin of the $Fe^{3+}$ ion, $S_x$, $S_y$, $S_z$ are the spin component along mutually perpendicular crystalline axes and D and E are the second-order crystal field terms with axial and rhombic symmetry, respectively. According to the literature [3, 4], if the $D$ value is much smaller than $g\mu_B H$ the spin Hamiltonian results in g-factor near 2 regardless of the E value, but for larger D values the effective g-factor can be around 4.3 or 6.0. Absorptions at $g_{eff} = 4.3$ and 6 are usually assigned to the isolated $Fe^{3+}$ ion located in distorted oxygen environments in sites with octahedral or tetrahedral symmetries, while the absorption at $g_{eff}=2.1$ is associated with strongly coupled $Fe^{3+}$ ions via super-exchange interaction.

## References

[1] C. Song, W. Xu, N. Liedienov, I. Fesych, R. Kulagin, Y. Beygelzimer, X. Zhang, Y. Han, Q. Li, B. Liu, A. Pashchenko, G. Levchenko. "Novel Multiferroic-Like Nanocomposite with High Pressure-Modulated Magnetic and Electric Properties." Advanced Functional Materials **32**, 2113022 (2022), https://doi.org/10.1002/adfm.202113022

[2] N. Iwamoto, Y.Makino, S. Kasahara**,** State of Fe3+ ion and Fe3+ −F− interaction in calcium fluorosilicate glasses, Journal of Non-Crystalline Solids, **55**, Issue 1, 113-124 (1983), https://doi.org/10.1016/0022-3093(83)90011-X

[3] N. O. Dantas, W. E.F. Ayta, A. C.A. Silva, Nilo F. Cano, A. F.R. Rodriguez, A. C. Oliveira, V. K. Garg, Paulo C. Morais, Magnetic and optical investigation of 40SiO2·30Na2O·1Al2O3·(29 − x)B2O3·xFe2O3 glass matrix, Solid State Sciences, **14**, Issue 8, 1169-1174 (2012), https://doi.org/10.1016/j.solidstatesciences.2012.05.033

[4] J. L. Ortiz-Quiñonez, et al, Easy synthesis of high-purity $BiFeO_3$ nanoparticles: new insights derived from the structural, optical and magnetic characterization, Inorg. Chem., **52**, 18, 10306–10317 (2013), https://doi.org/10.1021/ic400627c